\documentclass[aps,prb,twocolumn,superscriptaddress,]{revtex4-2}
\usepackage{cancel}
\usepackage[normalem]{ulem}
\usepackage{bm}
\usepackage{amsmath}
\usepackage{amsthm}
\usepackage{natbib}
\usepackage{amssymb}
\usepackage{graphicx}
\makeatletter
\usepackage{times}
\usepackage{comment}
\usepackage{graphicx}
\usepackage{feynmf}
\usepackage{tabularx}
\usepackage{amsmath}
\usepackage{amstext}
\usepackage{amssymb}
\usepackage[hidelinks,colorlinks=true,urlcolor=blue,linkcolor=blue,citecolor=blue]{hyperref}
\usepackage{longtable,booktabs}
\usepackage{url}
\usepackage{subfigure}
\usepackage{amsbsy}
\usepackage{amsthm}
\usepackage{bm}
\usepackage{cleveref}
\usepackage{mathrsfs}
\usepackage{amsfonts}
\usepackage{amsbsy}
\usepackage{color}
\usepackage{nicematrix} 
\usepackage{physics} 

\def\bphi{{\boldsymbol \phi}}
\def\brho{{\boldsymbol \brho}}
\def\bk{{\boldsymbol k}}
\def\bd{{\boldsymbol d}}
\def\bc{{\boldsymbol c}}
\def\bp{{\boldsymbol p}}
\def\bq{{\boldsymbol q}}
\def\br{\boldsymbol{r}}

\def\bs{{\boldsymbol s}}
\def\bx{{\boldsymbol x}}
\def\by{{\boldsymbol y}}

\def\bPhi{{\boldsymbol \Phi}}

\def\bpsi{{\boldsymbol \psi}}
\def\bJ{{\boldsymbol J}}
\def\bL{{\boldsymbol L}}
\def\bF{{\boldsymbol F}}

\def\bM{{\boldsymbol M}}

\def\bB{{\boldsymbol B}}
\def\bA{{\boldsymbol A}}

\def\la{\langle}
\def\calPT{\mathcal{PT}}

\def\calT{\mathcal{T}}

\def\calP{\mathcal{P}}

\def\calO{\mathcal{O}}

\def\pa{\partial}
\def\nn{\nonumber}

\def\la{\langle}
\def\ra{\rangle}
\newcommand\prts[1]{{{\scriptscriptstyle (}#1{\scriptscriptstyle )}}}
\newcommand\ua{\prts{a}}
\newcommand\ub{\prts{\raisebox{-0.08em}{$\scriptstyle b$}}}

\begin{document}
\title{Generalized Onsager reciprocal relations of charge and spin transport}
\author{Guan-Hua Huang}
\altaffiliation{These authors contributed equally to this work.}
\affiliation{Hefei National Laboratory, Hefei 230088, China}
\affiliation{Quantum Science Center of Guangdong-Hong Kong-Macao Greater Bay Area (Guangdong), Shenzhen 508045, China}
\author{Hui Tang}
\altaffiliation{These authors contributed equally to this work.}
\affiliation{Shenzhen Institute for Quantum Science and Engineering, Southern University of Science and Technology, Shenzhen 518055, China}
\author{Shizhong Zhang}
\affiliation{Department of Physics and HKU-UCAS Joint Institute for Theoretical and Computational Physics at Hong Kong, University of Hong Kong, Hong Kong, China}
\author{Zhongbo Yan}
\email{yanzhb5@mail.sysu.edu.cn}
\affiliation{Guangdong Provincial Key Laboratory of Magnetoelectric Physics and Devices, School of Physics, Sun Yat-Sen University, Guangzhou 510275, China}
\author{Zhigang Wu}
\email{wuzhigang@quantumsc.cn}
\affiliation{Quantum Science Center of Guangdong-Hong Kong-Macao Greater Bay Area (Guangdong), Shenzhen 508045, China}

\date{\today }
\begin{abstract}
In spin-orbit-coupled systems the charge and spin transport are generally coupled to each other, namely a charge current will induce a spin current and vice versa. In the presence of time-reversal symmetry $\calT$, the cross-coupling transport coefficients describing how one process affects the other are constrained by the famous Onsager reciprocal relations. In this paper, we generalize the Onsager reciprocal relations of charge and spin transport to systems that break the time-reversal symmetry but preserve a combined symmetry of $\calT$ and some other symmetry operation $\calO$. We show that the symmetry or antisymmetry of the cross-coupling transport coefficients remains in place provided that the operator $\calO$ meets certain conditions. Among many candidate systems where our generalized Onsager relations apply, we focus on a conceptually simple and experimentally realized model in cold atomic systems for explicit demonstration and use these relations to predict highly nontrivial transport phenomena that can be readily verified experimentally.  
\end{abstract}
\maketitle
\section{Introduction}
The Onsager reciprocal relations~\cite{OnsagerI,OnsagerII}, which reveal a fundamental symmetry or antisymmetry between the cross-coupling transport coefficients in time-reversal symmetric systems, are among the most important physical laws governing irreversible processes. When a system is perturbed slightly from equilibrium by a set of forces $F_j$ that are conjugate to generalized coordinates $x_j$, the induced current $J_i \equiv d x_i/d t$ corresponding to $x_i$ can be expressed  as $J_i = \sum_j L_{ij} F_j$ by linear response theory. The Onsager reciprocal relations then dictate that $L_{ij} = \pm L_{ji}$ in the presence of time-reversal symmetry, where the $+$($-$) sign arises if the product $x_ix_j$ is time-reversal even (odd)~\cite{Casimir}.  Connecting microscopic reversibility to thermodynamically irreversible processes, Onsager relations have provided deep insights into a wide range of transport phenomena~\cite{Miller1960,SHARIPOV1994457,10.1063/1.1782391,PhysRevE.69.016306,Jacquod2012,Gorini2012,Toth2003,Bhattacharya2014,PhysRevLett.129.238002,Pan2022,Guo2023,Tsang_2025}, including thermoelectric phenomena, superfluidity, and spintronics.  

Indeed, the establishment of Onsager reciprocal relations has played a pivotal role in addressing one of the long-debatted problems in spintronics, namely the proper definition of the spin current~\cite{Shi2006,Zhang2026,PhysRevB.98.081401,PhysRevB.102.125138,PhysRevB.104.L241411,tamaya2024}. Conventionally, the single-particle spin current operator is defined as $  \hat{\bm j}_{{\rm conv}}^{\rm s}= \frac{1}{2}\{ \hat s_z,\hat {\bm j}^{\rm c} \}$~\cite{Sinova2004,Sinova2015}, where $\hat s_z = \frac{1}{2}\sigma_z$ is the spin operator and $\hat{\bm j}^{\rm c}  = d\hat \br/dt$ is the single-particle charge (or mass) current operator. A central deficiency of this definition is that it does not correspond to any generalized coordinate and thus no conjugate force exists to generate this spin current, which in turn makes it impossible to establish the Onsager reciprocal relations in coupled charge and spin transport. To remedy this, Shi {\it et al.}~\cite{Shi2006} observed that the appropriate generalized coordinate for spin transport is the spin displacement operator $\hat\br \hat s_z$ and so the spin current can be defined as the corresponding total flux $\hat{\bm j}^{\rm s} = d (\hat \br \hat s_z )/dt$, in much the same way that the charge current is defined as the total flux of the charge displacement $\hat \br$. Thus, in response to the external potential $\delta V = -\bF^{\rm c} \cdot \hat \br  - \bF^{\rm s} \cdot (\hat \br \hat s_z)$, the charge current $\bJ^{\rm c}$ and the spin current  $\bJ^{\rm s}$ are determined by linear response theory as  
\begin{align}
J^{\rm c}_{\mu} &= \sum_\nu \left ( \sigma^{\rm cs}_{\mu\nu}F^{\rm s}_{\nu} +\sigma^{\rm cc}_{\mu\nu}F^{\rm c}_{\nu} \right ); \\
J^{\rm s}_{\mu} &= \sum_\nu \left (\sigma^{\rm ss}_{\mu\nu}F^{\rm s}_{\nu} +\sigma^{\rm sc}_{\mu\nu} F^{\rm c}_{\nu}\right ),
\end{align}
where $\bF^{\rm c}$ and $\bF^{\rm s}$ are forces conjugate to $\hat \br$ and $ \hat \br \hat s_z$ respectively, and $\{\sigma^{\rm ss}_{\mu\nu},\sigma^{\rm cc}_{\mu\nu},\sigma^{\rm sc}_{\mu\nu},\sigma^{\rm cs}_{\mu\nu}\}$ are a set of dc transport tensor coefficients~\footnote{To be specific, we focus on the components of the spin conductivity tensor describing the transport of spin aligned along $z$-direction in response to forces in the $xy$-plane.  }.
Assuming time-reversal symmetry, the Onsager reciprocal relations between the charge-spin cross-coupling transport coefficients are then~\cite{Shi2006} 
\begin{align}
\sigma^{\rm sc}_{\mu\nu}= -\sigma^{\rm cs}_{\nu\mu},
\label{onsager}
\end{align}
where the antisymmetry comes from the fact that the product $\hat r_\mu (\hat r_\nu \hat s_z)$ is time-reversal odd. This alternative definition of spin current and the associated Onsager relations have been influential for several reasons~\cite{Sugimoto2006,Liu2023,Cullen2023,MaCullen2024}. For example, the spin Hall conductivity $\sigma^{\rm sc}_{xy}$, traditionally challenging to measure, can now be deduced from the electric current induced by a spin force in light of the Onsager relations in Eq.~(\ref{onsager}). 

The Onsager reciprocal relations are fundamentally predicated on the time-reversal symmetry, yet this symmetry is broken by magnetism in many systems actively studied for spintronics~\cite{Jungwirth2016,smejkal2018,Baltz2018,DalDin2024,Rimmler2025}. Therefore, to what extent these relations can still be established in such systems remains a pressing open question~\cite{Seemann2015,Luo2020,zhou2025}.  The critical importance of resolving this question is exemplified by 
recent contradictory findings in antiferromagnet YbMnBi$_{2}$, where independent studies have reported both the preservation 
and the violation of the Onsager reciprocal relations regarding the anomalous electric and thermoelectric conductivity tensors~\cite{Pan2022,Guo2023}.

 In this paper, we generalize the Onsager reciprocal relations of charge and spin transport to systems that break the time-reversal symmetry $\calT$ but preserve a combined symmetry of $\calT$ and some other symmetry operation $\calO$.  If the symmetry operation $\calO$ meets certain conditions, reciprocal relations analogous to that in Eq.~(\ref{onsager}) can still be established. In Sec.~\ref{secGOR}, we derive these relations based on  symmetry analysis and classify the appropriate  $\calO$ symmetry operations into three categories according to how the charge and spin displacement transform under the $\calO$ operation. In Sec.~\ref{QG} we illustrate all three categories of $\calO$ operations and the corresponding generalized Onsager relations using a single model of 2D spin-orbit-coupled quantum gases recently realized in experiments. Furthermore, we explicitly verify these  relations by evaluating the charge-spin cross-coupling transport coefficients. Due to the experimental relevance of the model, the generalized Onsager relations can be experimentally tested. In Sec.~\ref{proposal}, we propose an experimental scheme for this purpose and our numerical simulations show that these relations lead to unexpected charge-spin transport phenomena. All these results are summarized in Sec.~\ref{conclusion}.

\section{Generalized Onsager relations of charge and spin transport}
\label{secGOR}
In this section we show that for systems that preserve a composite $\calO\calT$ symmetry,  relations similar to Eq.~(\ref{onsager}) still hold  provided that $\hat r_\mu$ and $\hat s_z$ are invariant or change sign under operation $\calO$, i.e., 
\begin{align}
 \calO^{-1}\hat r_\mu \calO = (-1)^{I_\mu} \hat r_\mu; \quad \calO^{-1}\hat s_z \calO = (-1)^{I_s} \hat s_z 
 \label{O},
 \end{align} 
 where the indices $I_\mu$ and $I_s$ take the value of $0$ or $1$.   
We begin with the Kubo formulas for the charge-spin cross-coupling transport coefficients at finite frequency
\begin{align}
\label{sigmakubo1}
\sigma^{\rm sc}_{\mu\nu} (\omega) &= i\int_{0}^\infty d t \la     [  {\hat J}^{\rm s}_{\mu} (t) , \hat R^{\rm c}_{\nu} (0)  ]  \ra  e^{i\omega t };  \\
\sigma^{\rm cs}_{\nu\mu} (\omega) &= i \int_{0}^\infty d t \la   [  {\hat J}^{\rm c}_{\nu} (t) , \hat R^{\rm s}_{\mu} (0)  ]  \ra  e^{i\omega t}.
\label{sigmakubo2}
\end{align}
Here $\la \cdots \ra$ stands for a statistical average ${\rm Tr}(\hat \varrho \cdots)$ with respect to the density matrix $\hat \varrho$ in general which reduces to the expectation value with respect to the ground state at zero temperature. Here $\hat R^{\rm c}_{\nu} = \sum_j \hat r_{j,\nu}$ and $\hat R^{\rm s}_{\mu} = \sum_j \hat r_{j,\mu}\hat s_{j,z}$ are the many-body charge and spin displacement operators respectively, and $\hat J^{\rm c}_{\nu}(t) = d \hat R^{\rm c}_{\nu}(t)/ dt$  and $\hat J^{\rm s}_{\mu}(t) = d \hat R^{\rm s}_{\mu}(t)/ dt$ are the corresponding charge and spin current operators.  As usual, the Heisenberg operators such as $\hat J^{\rm s}_{\mu}(t)$ is defined as $\hat J^{\rm s}_{\mu} (t) = e^{i\hat H t} \hat J^{\rm s}_{\mu}(0) e^{-i\hat H t}$, where $\hat H $ is the Hamiltonian of the system ($\hbar  = 1$ throughout the paper). 

We now derive the generalized Onsager relations for systems described by a density matrix $\hat\varrho$ that respects the $\calO\calT$ symmetry, i.e., $[\hat \varrho, \calO\calT ] =0$. Let $\{|\Phi_n\ra\}$ be a complete set of eigenstates of the Hamiltonian. Denoting $|\tilde \Phi_n\ra = \calO\calT |\Phi_n\ra$ and keeping in mind that $\calO\calT$ is an antiunitary operator, we have 
\begin{align}
\la \Phi_n|\hat \varrho \hat A^\dag |\Phi_n\ra = \la \tilde\Phi_n|\calO\calT\hat A \hat\varrho^\dag (\calO\calT)^{-1}| \tilde\Phi_n\ra
\end{align}
 for any operator $\hat A$. Since $\hat \varrho$ is Hermitian and $\hat \varrho = \calO\calT\hat \varrho (\calO\calT)^{-1}$ due to the $\calO\calT$ symmetry, we find 
 \begin{align}
 \la \Phi_n|\hat \varrho \hat A^\dag |\Phi_n\ra =  \la \tilde \Phi_n|\calO\calT \hat A (\calO\calT )^{-1}\hat \varrho |\tilde \Phi_n\ra.
 \label{Phin}
 \end{align}
  Because $\{|\tilde\Phi_n\ra\}$ also form a complete set of states, summation of both sides of Eq.~(\ref{Phin}) over all the states leads to 
\begin{align}
{ \rm Tr}(\hat \varrho \hat A^\dag )  = { \rm Tr}\big [\hat \varrho \,\calO\calT\hat A (\calO\calT)^{-1} \big ].
\end{align}
 Now we take $\hat A = \hat R^{\rm c}_{\nu}(0) \hat R^{\rm s}_{\mu}(t)$ and first observe that
\begin{align}
\calO\calT\hat R^{\rm c}_{\nu}(0) (\calO\calT)^{-1} = (-1)^{I_\nu}\hat R^{\rm c}_{\nu}(0) 
\label{ROT}
\end{align}
 due to the $\calO$ operation in Eq.~(\ref{O}) and the time-reversal invariance of $\hat R^{\rm c}_{\nu}(0)$. Furthermore, since the antiunitary operator $\calO\calT$ commutes with $\hat H$ we find that
 \begin{align}
 \calO\calT\hat R^{\rm s}_{\mu}(t) (\calO\calT)^{-1} &= e^{-i\hat H t}\calO\calT\hat R^{\rm s}_{\mu}(0) (\calO\calT)^{-1}  e^{i\hat H t} \nn \\
 & = (-1)^{I_\mu+I_s+1}\hat R^{\rm s}_{\mu}(-t),
 \end{align}
 where we again used Eq.~(\ref{O}) and the fact that the spin is time-reversal odd in arriving at the second line. Putting these together, we find that 
\begin{align}
\la \hat R^{\rm s}_{\mu} (t) \hat R^{\rm c}_{\nu} (0) \ra = (-1)^{I_{\mu\nu}}\la \hat R^{\rm c}_{\nu} (t) \hat R^{\rm s}_{\mu} (0) \ra,
\end{align}
where the time-translation symmetry is used for the expectation value and $I_{\mu\nu} \equiv I_\mu + I_\nu +I_s+1$. Taking the derivative on both sides further yields 
\begin{align}
\la   {\hat J}^{\rm s}_{\mu} (t) \hat R^{\rm c}_{\nu} (0) \ra = (-1)^{I_{\mu\nu}}\la {\hat J}^{\rm c}_{\nu} (t) \hat R^{\rm s}_{\mu} (0)  \ra.
\label{JR}
\end{align}
Lastly, substituting Eq.~(\ref{JR}) into Eqs.~(\ref{sigmakubo1})-(\ref{sigmakubo2}) we arrive at the generalized Onsager reciprocal relations of charge and spin transport
\begin{align}
\sigma^{\rm sc}_{\mu\nu}(\omega) = (-1)^{I_{\mu\nu}}\sigma^{\rm cs}_{\nu\mu}(\omega).
\label{gonsager}
\end{align}

Several comments are now in order. First, our generalized Onsager relations should be distinguished from the Onsager-Casimir relations~\cite{Casimir} which are extensions of the Onsager relations to account for time-reversal symmetry breaking by an external magnetic field (see Appendix~\ref{AppA}). Second, we recover the usual Onsager relations in Eq.~(\ref{onsager}) in the presence of time-reversal symmetry by setting $\calO$ as an identity operator and taking the zero-frequency limit.  Third, the conditions for operation $\calO$ in Eq.~(\ref{O}) can actually be somewhat relaxed. It is clear that the transformation of the commutators in Eqs.~(\ref{sigmakubo1})-(\ref{sigmakubo2}) under $\calO$ remains the same even if the right-hand sides of the equations in Eq.~(\ref{O}) include an additional finite constant. This means that any $\calO$ operation leaving $\hat r_\mu$ and $\hat s_z$ invariant or with a changed sign \emph{up to a constant} still leads to the generalized Onsager relations in Eq.~(\ref{gonsager}).  Fourth,  suitable $\calO$ operations fall into three categories: (i) pure spatial operations such as spatial inversion $\calP$, $\pi$-rotation about any Cartesian axis and lattice translation; (ii) pure spin operations such as any spin rotation around $z$-axis and $\pi$-rotation around $x$ or $y$-axis; and (iii) space-spin operations including any combination of those in (i) and (ii), as well as mirror and glide symmetries. 
Lastly, in cases where the system has multiple $\calO\calT$ symmetries, we must examine the generalized Onsager relations corresponding to each and every $\calO$ operation. If two $\calO$ operations lead to contradictory relations with respect to $\sigma^{\rm sc}_{\mu\nu}(\omega)$ and $\sigma^{\rm cs}_{\nu\mu}(\omega)$, we can immediately draw the conclusion from Eq.~(\ref{gonsager}) that both must vanish identically. 

 The generalized Onsager reciprocal relations derived above are expected to find broad application, 
particularly in antiferromagnets where zero net magnetization is enforced by specific forms
of $\mathcal{O}\mathcal{T}$ symmetry~\cite{LDYuan2021,Libor2022AMa}. Notably, 
many experimentally realized antiferromagnets (see, e.g., Table I in Ref.~\cite{Hayami2022}), 
including those with nontrivial topological properties~\cite{Tang2016},  
are found to possess $\mathcal{P}\mathcal{T}$ symmetry as well as other combined symmetries 
of $\mathcal{T}$ and translation, rotation or mirror reflection operations. While 
candidate electronic materials are abundant, the simultaneous measurements of both
$\sigma^{\rm sc}_{\mu\nu}(\omega)$ and $\sigma^{\rm cs}_{\nu\mu}(\omega)$ in these systems 
remain experimentally challenging. Furthermore, numerical verification of these relations in any of these materials either involve {\it ab initio} calculations that are computationally costly or are based on model Hamiltonians that are necessarily simplifications of the real systems. For these reasons, quantum gases---with their significantly 
greater flexibility---provide an ideal platform to demonstrate these generalized Onsager relations as a proof of principle.

\section{Illustration of Generalized Onsager relations in quantum gases}
\label{QG}
\begin{figure}[tb]
\begin{centering}
\includegraphics[width=6.5cm]{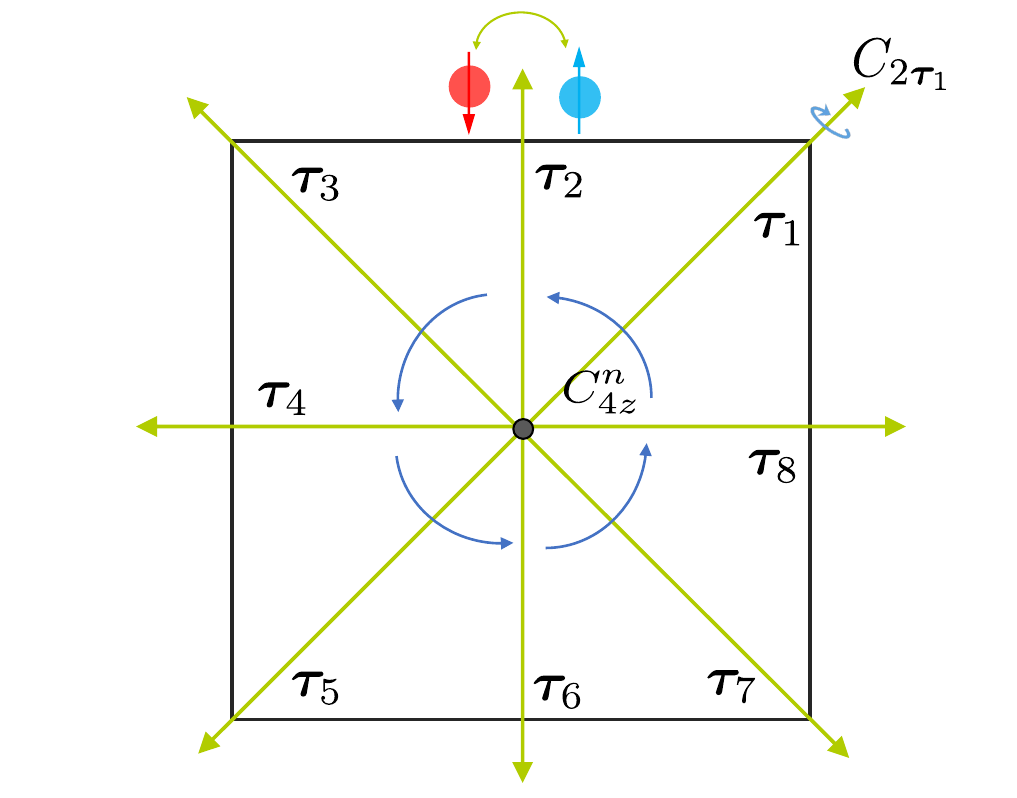}
\par\end{centering}
\caption{Illustration of the symmetry operations in the $\tilde D_4$ group. }
\label{fig:D4}
\end{figure}
To showcase the versatility of quantum gas systems, we now use a single model of 2D spin-orbit-coupled atomic gases to illustrate all three categories of $\calO$ operations and the resulting generalized Onsager relations. For simplicity, we will focus mostly on the case of zero temperature for the rest of the discussion. Realized experimentally with both fermionic species $^{87}$Sr~\cite{Liang2023} and bosonic species $^{87}$Rb~\cite{Sun2018}, the Hamiltonian of this model is
\begin{align}
\hat H = \sum_{\sigma\sigma'}\int d\br \left (\hat\psi^\dag_{\sigma}\hat h_{\sigma\sigma'} \hat \psi_{\sigma'}+\frac{g_{\sigma\sigma'}}{2}\hat\psi^\dag_{\sigma}\hat\psi^\dag_{\sigma'} \hat \psi_{\sigma'}\hat \psi_{\sigma}\right ),
\label{MBH}
\end{align}
where $\hat\psi^\dag_{\sigma}(\br)$ ($\sigma = \uparrow,\downarrow$) creates a particle of spin $\sigma$, $g_{\sigma\sigma'}$ are the $s$-wave atomic interaction constants and $\hat h$ is the single-particle Hamiltonian 
 \begin{align}
	\label{h0}
	\hat h=\left[{ \hat \bp^2}/{2m}  +V_{\text{latt}}(\hat \br) \right ]  +V_{\rm R 1}(\hat\br) \hat s_x + V_{\rm R2}(\hat\br)\hat s_y.
\end{align}
Here $V_{\text{latt}}(\hat \br)=V_0(\cos^2k_{L}\hat r_x+\cos^2k_{L}\hat r_y)$ is a 2D square optical lattice potential with potential depth $V_0$ and lattice separation $\pi/k_L$; the spin-orbit coupling (SOC) is simulated using two Raman lattice potentials $ V_{\rm R1}(\br)=2M_0\sin (k_{L}\hat r_x)\cos (k_{L}\hat r_y)$ and $V_{\rm R2}(\hat \br) = 2M_0\sin (k_L\hat r_y)\cos (k_{L}\hat r_x)$, which couple the spatial degrees of freedom to the spin components $\hat s_x = \frac{1}{2}\sigma_{x}$ and $\hat s_y =\frac{1}{2} \sigma_{y}$. 

A salient property of this model is that it has a high degree of symmetry~\cite{Huang2021,Chen2023} including the $\calP\calT$ symmetry and the modified dihedral symmetry $\tilde D_4$. The latter is a double group whose $16$ elements are spin-space rotations along various symmetry axes of the square lattice, i.e.,  four-fold rotation around the $z$-aixs
\begin{align}
C_{4z}^n = e^{-i\frac{n\pi}{2}( \hat s_z +\hat L_z)} 
\end{align} 
and  two-fold rotation around the in-plane vector $\boldsymbol \tau_n$ 
\begin{align}
 C_{2 {\boldsymbol \tau}_n} = e^{-i \pi  {\boldsymbol \tau}_n \cdot (\hat \bs +\hat\bL)},
\end{align}
where ${\boldsymbol \tau}_n = \cos(n\pi/4)\hat\bx + \sin(n\pi/4) \hat\by$ are in-plane unit vectors with $n = 1,2,\cdots,8$ [see Fig.~(\ref{fig:D4})]. It can be verified that the single-particle Hamiltonian in Eq.~(\ref{h0}) and the usual atomic interactions both commute with $\calPT$ and all elements of $\tilde D_4$, and naturally with any combination of them. Thus the Hamiltonian has the $\calO\calT$ symmetry where $\calO \in \calP\tilde D_4$. Among the $16$ elements of $\calP\tilde D_4$,  four operators have the properties in Eq.~(\ref{O}), which are $\calP$, $C^2_{4z}\mathcal{P}=e^{-i\pi\hat s_z}$, $C_{2{\boldsymbol \tau}_2}\mathcal{P} = e^{-i\pi(\hat s_y+\hat L_x)}$ and $C_{2{\boldsymbol \tau}_4} \mathcal{P} = e^{i\pi(\hat s_x+\hat L_y)}$. It's clear that $\calP$ and $C^2_{4z}\calP$, respectively, belong to the category (i) and (ii) symmetry operations previously defined, while $C_{2{\boldsymbol \tau}_2}\mathcal{P} $ and $C_{2{\boldsymbol \tau}_4} \mathcal{P} $ belong to category (iii). We thus see that the four $\calO$ operations indeed encompass all three categories. The transformations of $\hat r_\mu$ and $\hat s_z$ under these operations are shown in the top section of Tab.~\ref{threesys}.

\begin{table}[tb]
\caption{Generalized Onsager relations in quantum gases}
\renewcommand\arraystretch{1.3}
\begin{NiceTabular}{@{}cccccc@{}}[columns-width = 1.4cm]
\toprule
$\mathcal{O}$ & $\mathcal{P}$ & $C^2_{4z}\mathcal{P}$ & $C_{2{\boldsymbol \tau}_2}\mathcal{P}$ & $C_{2{\boldsymbol \tau}_4}\mathcal{P}$\\
\hline $I_{x}$ & $1$ & $0$ & $0$ & $1$\\
$I_{y}$ & $1$ & $0$ & $1$ & $0$\\
$I_{s}$ & $0$ & $0$ & $1$ & $1$\\
\hline 
$(-1)^{I_{xx}}$ & $-$ & $-$ & $+$ & $+$\\
$(-1)^{I_{xy}}$ & $-$ & $-$ & $-$ & $-$\\
\hline 
Fermi & \checkmark & \checkmark & \checkmark & \checkmark \\
Bose (phase I) & $\times$ & $\times$ & \checkmark & \checkmark\\
Bose (phase II) & $\times$ & \checkmark & $\times$ & $\times$ \\
\bottomrule
\end{NiceTabular}
\vspace{-1em}
\label{threesys}
\end{table}

The ground states of this model, however, do not necessarily preserve all the symmetries of the Hamiltonian. Depending on particle statistics as well as atomic interactions, this Hamiltonian supports three types of ground states, each characterized by a different set of $\calO\calT$ symmetries.
For spin-$\frac{1}{2}$ fermions with weak repulsive interactions, the ground state does not break any symmetry of the Hamiltonian and so the $\calO\calT$ symmetries with previously mentioned four $\calO$ operations are all present. Each of the $\calO\calT$ symmetry leads to a set of generalized Onsager relations, which might not be the same. As can be seen in the middle section of Tab.~\ref{threesys}, the generalized Onsager relation for the longitudinal coefficients obtained from $\calP$ and $C^2_{4z} \calP$ are 
\begin{align}
\sigma^{\rm sc}_{xx}(\omega) = -\sigma^{\rm cs}_{xx}(\omega) 
\end{align}
while those from $C_{2{\boldsymbol \tau}_2}\calP$ and $C_{2{\boldsymbol \tau}_4}\calP$ are
\begin{align}
\sigma^{\rm sc}_{xx}(\omega) = \sigma^{\rm cs}_{xx}(\omega).
\end{align}
Thus we can infer immediately that 
\begin{align}
\sigma^{\rm sc}_{xx}(\omega) = \sigma^{\rm cs}_{xx}(\omega) =0
\label{xxzero}
\end{align} 
for the Fermi gas. However, for the transverse coefficients all four $\calO$ operations lead to the same generalized Onsager relation 
\begin{align}
\sigma_{xy}^{\text{sc}}(\omega)=-\sigma_{yx}^{\text{cs}}(\omega).
\label{fermixy}
\end{align}

The situation is different for the two-component Bose gas where the ground state necessarily breaks some symmetries of the Hamiltonian due to Bose-Einstein condensation. Here we focus on the case of  $g_{\uparrow\uparrow} = g_{\downarrow\downarrow} > g_{\uparrow\downarrow}$ as considered in experiments. Both theory and experiments have shown that the interplay of SOC and atomic interaction lead to two phases distinguished by the magnetization, namely a perpendicularly magnetized phase at large SOC strengths (phase I) and an in-plane magnetized phase at small SOC strengths (phase II) [see Fig.~\ref{fig:phasediagram} in Appendix~\ref{tcbose}]. These two phases can also be differentiated by which $\calO \calT$ symmetries they have retained~\cite{Chen2023} (see also Appendix~\ref{tcbose}). For the perpendicularly magnetized state (phase I), the condensate wave function $\bPhi_{\rm I}(\br)$ is invariant with respect to both $C_{2{\boldsymbol \tau}_2}\calPT $ and $C_{2{\boldsymbol \tau}_4} \calP \calT$. According to the middle section of Tab.~\ref{threesys}, both symmetry operations lead to the same set of generalized Onsager relations 
\begin{align}
 \sigma_{xx}^{\text{sc}}(\omega)=\sigma_{xx}^{\text{cs}}(\omega); \quad \sigma_{xy}^{\text{sc}}(\omega)=-\sigma_{yx}^{\text{cs}}(\omega).
 \label{OphaseI}
 \end{align}
  It's interesting to note that the generalized Onsager relation in phase I is symmetric for longitudinal dynamics but antisymmetric for transverse dynamics.  
In contrast, the condensate wave function $\bPhi_{\rm II}(\br)$ for the in-plane magnetized state (phase II) only has the single combined symmetry of $C^2_{4z} \calP\calT$, which results in 
  \begin{align}
 \sigma_{xx}^{\text{sc}}(\omega)=-\sigma_{xx}^{\text{cs}}(\omega); \quad \sigma_{xy}^{\text{sc}}(\omega)=-\sigma_{yx}^{\text{cs}}(\omega).
 \label{OphaseII}
 \end{align}
   These three ground states and the $\calO\calT$ symmetries that they preserve are summarized in the lower section of Tab.~\ref{threesys}.

 \section{Verification of Generalized Onsager relations in quantum gases}
\label{QGv}
In this section, we explicitly verify the generalized Onsager relations established in the last section by calculating the transport tensors without invoking the symmetry analysis. Unfortunately, the expressions in Eqs.~(\ref{sigmakubo1})-(\ref{sigmakubo2}) are inconvenient for a lattice system with periodic boundary conditions due to the fact that evaluating the matrix elements of the charge displacement operator $\hat R^{\rm c}_\mu$ between the Bloch states involves some subtle difficulty~\cite{MaCullen2024}. Instead, the following equivalent formulas can be used (see Appendix~\ref{AppB})
\begin{align}
\label{sigmakubo3}
\sigma^{\rm sc}_{\mu\nu}(\omega) &= \lim_{\bq\rightarrow 0}{\pa_{q_\mu}} \! \int_{0}^\infty \!  d t \la     [   \hat \rho^{\rm s}(\bq,t),  \hat j^{\rm c}_{\nu} (-\bq,0)  ]   \ra  e^{i\omega t }; \\ 
\sigma^{\rm cs}_{\nu\mu}(\omega) &= \lim_{\bq\rightarrow 0}{\pa_{q_\mu}} \!\int_{0}^\infty \! d t \la     [  \hat j^{\rm c}_{\nu} (\bq,t) , \hat \rho^{\rm s}(-\bq,0) ]   \ra  e^{i\omega t },
\label{sigmakubo4}
\end{align}
where 
\begin{align}
\hat \rho^{\rm s}(\bq) = \sum_j \hat s_{j,z} e^{-i\bq\cdot \hat \br_j}
\end{align}
 is the spin density and 
 \begin{align}
   \hat j^{\rm c}_{\nu}(\bq) = \frac{1}{2}\sum_j (\frac{d {\hat r}_{j,\nu}}{dt}  e^{-i\bq\cdot \hat\br_j} +e^{-i\bq\cdot \hat\br_j} \frac{d\hat r_{j,\nu} }{dt}  )
\end{align} is the charge current density.  As per the standard procedure, we first calculate the finite temperature spin-current correlation functions~\cite{mahan2013many},
\begin{align}
    \chi_{\rho^{\rm s},j_\nu^{\rm c}}(\bm q,\tau)= &-\langle\mathscr{T}\hat \rho^{\rm s}(\bm q,\tau)\hat j_{\nu}^{\rm c}(-\bm q,0)\rangle; \label{chi_rhoJ1} \\
        \chi_{j_\nu^{\rm c}, \rho^{\rm s}}(\bm q,\tau)= &-\langle\mathscr{T}\hat j_\nu^{\rm c}(\bm q,\tau)\hat \rho^{\rm s}(-\bm q,0)\rangle, \label{chi_rhoJ2}    
\end{align}
where $\tau\in [0,\beta)$ is the imaginary time, $\beta = 1/T$ is the inverse temperature and $\mathscr{T}$ represents the imaginary time ordering operation. Afterwards we perform analytic continuations of the Fourier components of these functions to the real frequency domain to obtain $\sigma^{\rm sc}_{\mu\nu}(\omega)$ and $\sigma^{\rm cs}_{\mu\nu}(\omega)$.
\begin{figure}[tb]
\begin{centering}
\includegraphics[width=8.6cm]{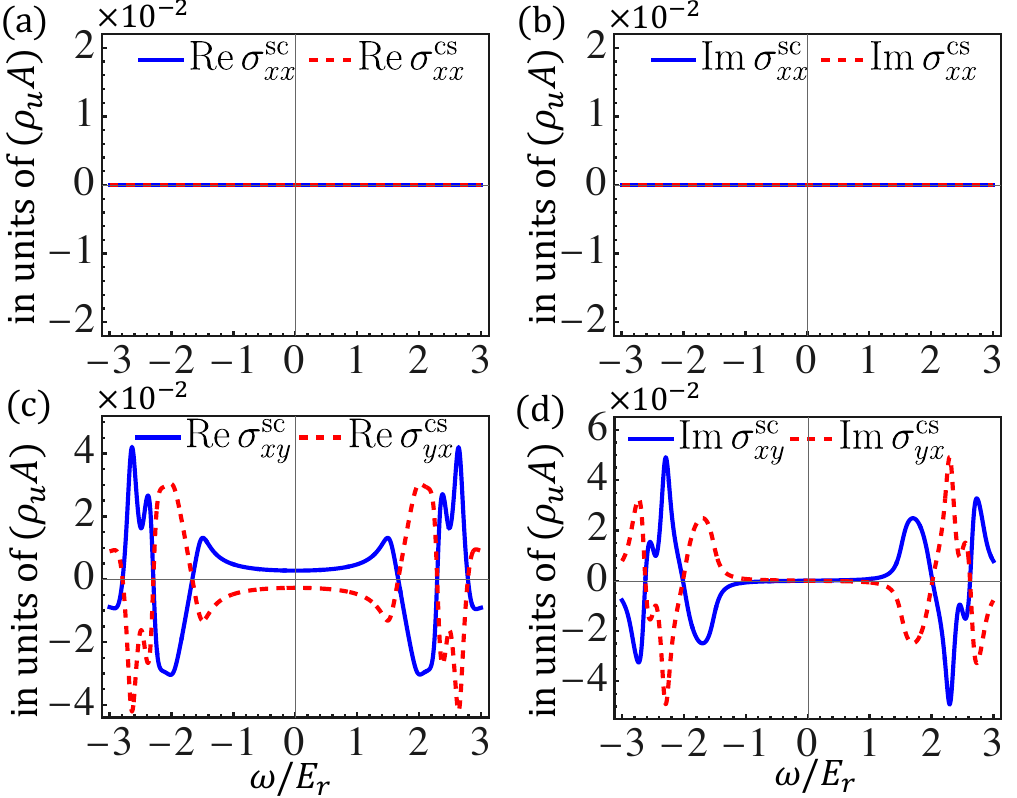}
\par\end{centering}
\caption{Verification of the generalized Onsager relations for the noninteracting Fermi
gas. The conductivity tensors are in units of $\rho_u A$ where $\rho_u = 2\rho \pi^2 /k_L^2$ is the atom number per unit cell and $A$ is the area of the system. The  parameters are $\rho_u=4$, $V_0=4.0E_r$ and $M_0=2.0E_r$; a spectral broadening  $\eta = 0.1E_r$ is used in plotting. }
\label{fig:GORfermi}
\end{figure}

For the degenerate Fermi gas, we restrict ourselves to the noninteracting case for simplicity. The  charge-spin correlation functions can be expressed in terms of the single-particle Green's function, which can be calculated straightforwardly (see details in Appendix~\ref{tcfermi}). As shown in Fig.~\ref{fig:GORfermi}(a) and (b), the longitudinal coefficients  $\sigma^{\rm cs}_{xx}(\omega)$ and  $\sigma^{\rm cs}_{xx}(\omega)$ calculated using Eqs.~(\ref{sigmakubo3})-(\ref{sigmakubo4}) vanish identically, validating the relation in Eq.~(\ref{xxzero}).  The transverse cross-coupling coefficients as shown in Fig.~\ref{fig:GORfermi}(c) and (d) are clearly antisymmetric, in perfect agreement with Eq.~(\ref{fermixy}).

\begin{figure}[tb]
\begin{centering}
\includegraphics[width=8.6cm]{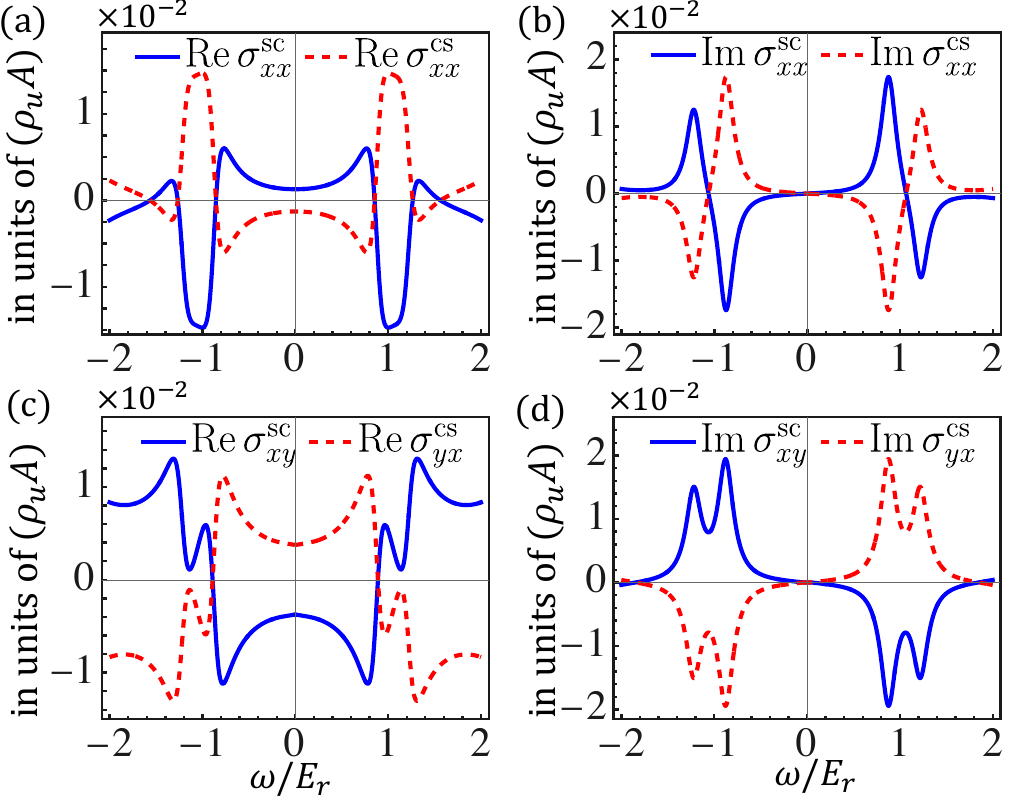}
\par\end{centering}
\caption{Verification of the generalized Onsager relations for the Bose
gas in the in-plane magnetized phase (phase II). Here the parameters are $V_0=4.0E_r$, $M_0=1.0E_r$, $\rho g_{\uparrow\uparrow}=0.25E_r$, $\rho g_{\uparrow\downarrow}=0.2E_r$  and the spectral broadening $\eta = 0.1E_r$. }
\label{fig:GORBII}
\end{figure}

For the Bose condensate, the calculation of the charge-spin transport coefficients are slightly more involved. Under the Bogoliubov approximation, the spin density and charge current density operators can be expressed in terms of the quasi-particle operators and the charge-spin correlation functions can be evaluated once the excitation spectrum is determined~\cite{Tang2025} (see details in Appendix~\ref{tcbose}).  Shown in Fig.~\ref{fig:GORBI} and Fig.~\ref{fig:GORBII} are $\sigma^{\rm sc}_{\mu\nu}(\omega)$ and $\sigma^{\rm cs}_{\nu\mu}(\omega)$ calculated using Eqs.~(\ref{sigmakubo3})-(\ref{sigmakubo4}) for the Bose gas in the perpendicularly magnetized phase (phase I) and in-plane magnetized phase (phase II) respectively. These results show perfect agreement with the generalized Onsager relations given in Eq.~(\ref{OphaseI}) and Eq.~(\ref{OphaseII}). 

\begin{figure}[tb]
\begin{centering}
\includegraphics[width=8.6cm]{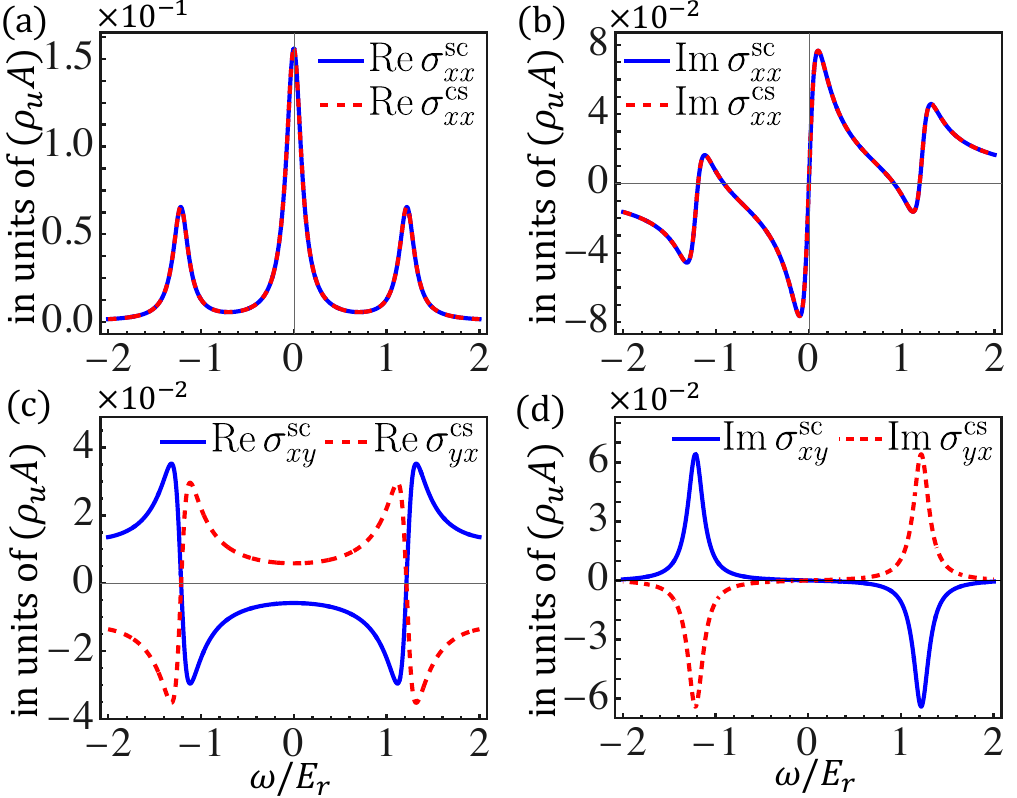}
\par\end{centering}
\caption{Verification of the generalized Onsager relations for the Bose gas in perpendicularly magnetized phase (phase I). Here $V_0 = 4E_r$, $M_0= 2.5E_r$, $\rho g_{\uparrow\uparrow} = 0.25 E_r $ and $ g_{\uparrow\downarrow} /g_{\uparrow\uparrow}= 0.9954$ (experimental ratio), where $\rho$ is the average atomic density and $E_r = k_L^2/2m$ is the recoil energy. A spectral broadening of $\eta  = 0.1E_r$ is again used. }
\label{fig:GORBI}
\end{figure}

\begin{figure}[b]
\begin{centering}
\includegraphics[width=8.6cm]{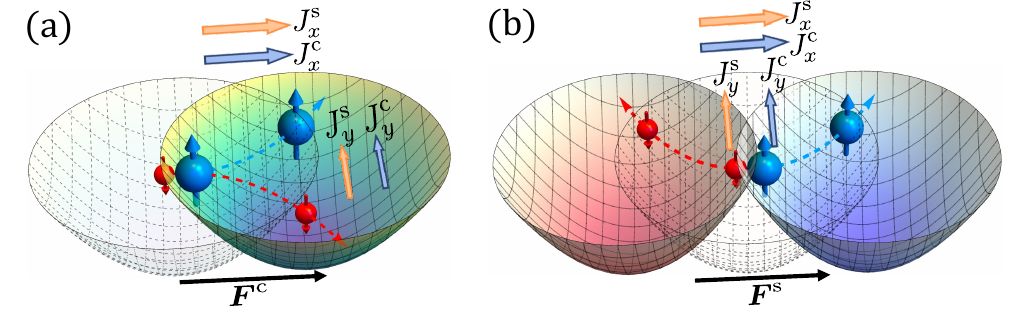}
\par\end{centering}
\caption{Illustration of two experimental quench protocols that can be used to generate a charge force [protocol (a)] and a spin force [protocol (b)] in harmonically trapped quantum gases.}
\label{fig:proposal}
\end{figure}
\section{Experimental proposal and simulations}
\label{proposal}
In this section, we discuss how to test the generalized Onsager relations experimentally in cold atomic systems confined in an additional harmonic trapping potential $V_{\rm tr}(\br) = m\omega_0^2 \br^2 /2$.  Such trapped atomic systems are uniquely suited for the experimental study of these relations for two reasons. First, the forces conjugate to charge and spin displacements can be easily generated and controlled in experiments.  For example, a spatially uniform charge force $\bF^{\rm c}(t) = \theta(t) m\omega_0^2 \bd $ can be produced by abruptly displacing the trapping potential a distance of $\bd$ at time $t = 0$, i.e., by quenching the trapping potential to $\tilde V_{\rm tr}(\br) = V_{\rm tr}(\br -\bd) $ [see Fig.~\ref{fig:proposal}(a)]. Similarly, a spin force $\bF^{\rm s}(t) = \theta(t) m\omega_0^2 \bd $ can be generated by replacing the trapping potential by two spin-selective potentials~\cite{PhysRevA.75.053612} separated by a distance of $2\bd$, i.e., by quenching the trapping potential to 
$\tilde V_{\rm tr} (\br)= V_{\rm tr}(\br -\bd)\ketbra{\uparrow}{\uparrow} + V_{\rm tr}(\br +\bd)\ketbra{\downarrow}{\downarrow} $ [see Fig.~\ref{fig:proposal}(b)]. Second, the coupled charge-spin dynamics can be conveniently probed by measuring the charge and spin displacements instead of the currents~\cite{Wu2015,Anderson2019}. According to the relation between the generalized coordinate and the currents, the spin displacement resulting from the quench protocol (a) in Fig.~\ref{fig:proposal}(a) is given by
\begin{align}
R^{{\rm s}\ua}_{\nu}(t) = \int_0^t dt' \int  d \omega\, \sigma^{\rm sc}_{\nu\mu}(\omega)  F^{\rm c}_{\mu}(\omega) e^{-i\omega t'}
\end{align}
and the charge displacement from the quench protocol (b) in Fig.~\ref{fig:proposal}(b) is 
\begin{align}
R^{{\rm c}\ub}_{\mu} (t) =  \int_0^t dt' \int d \omega\, \sigma^{\rm cs}_{\mu\nu}(\omega)  F^{\rm s}_{\nu}(\omega) e^{-i\omega t'},
\end{align}
where $\bF^{\rm c\prts{s}}(\omega) = \int dt \bF^{\rm c\prts{s}}(t) e^{i\omega t}$. Assuming that both quenches are along the $x$-direction, we then have $\bF^{\rm c}(t) = \bF^{\rm s} (t) = \theta (t) m\omega_0^2 d \hat \bx$. Thus, the generalized Onsager relations for the Bose gas in phase I can be tested by examining whether the measured spin and charge displacements from these two protocols satisfy the following relations 
\begin{align}
R^{{\rm s}\ua}_{x}(t) =  R^{{\rm c}\ub}_{x}(t); \quad R^{{\rm s}\ua}_{ y} (t) = - R^{{\rm c}\ub}_{y} (t).
\label{Rrelations}
\end{align}

 \begin{figure}[t]
\begin{centering}
\includegraphics[width=8.6cm]{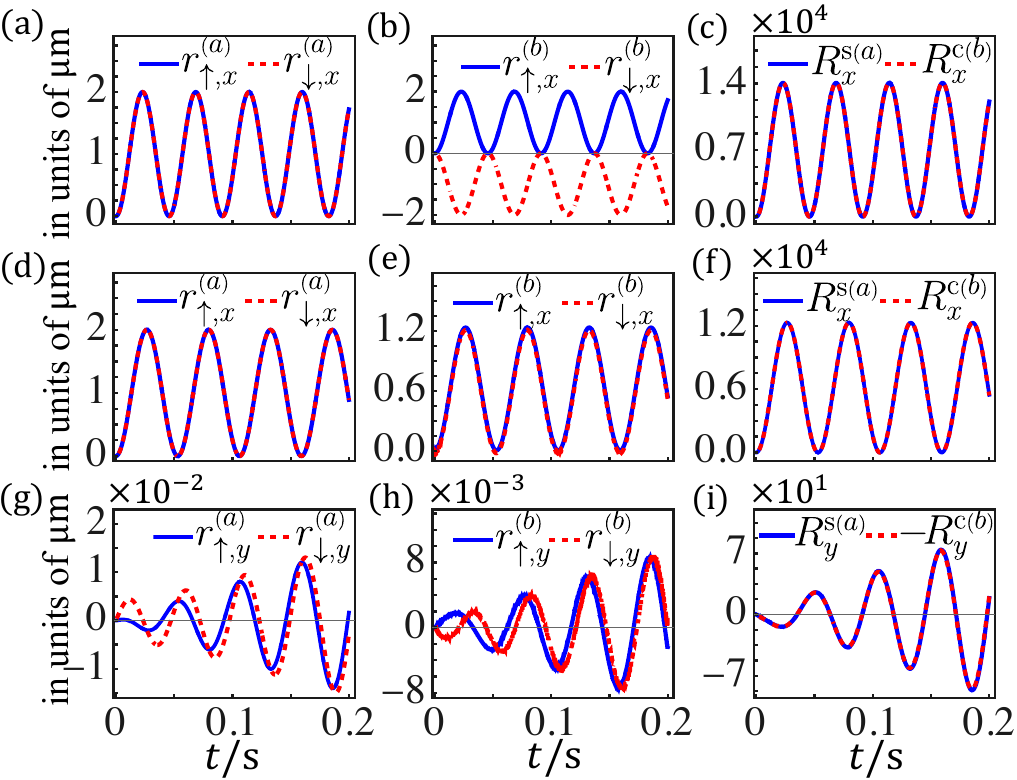}
\par\end{centering}
\caption{The c.m. dynamics of the Bose gas in phase I during quench protocol (a) (the first column) and (b) (the second column) at zero temperature ($\gamma = 0$). The first and the second row present the longitudinal dynamics of the independent two-component system ($M_0 = g_{\uparrow\downarrow } = 0$)  and the spin-orbit-coupled system respectively. The third row presents the transverse dynamics of the latter system. Here the trap frequency is $\omega_0 = 2\pi\times 27$Hz, the trap displacement is $d = 1\mu$m, the total number of atoms is $N = 10^4$ and $g_{{\uparrow\uparrow}} =1.946\times10^{-11} $Hz m$^2$. The rest of the  parameters are the same as those in Fig.~\ref{fig:GORBI}. }
\label{fig:GP1}
\end{figure}

\begin{figure}[b]
\begin{centering}
\includegraphics[width=8.6cm]{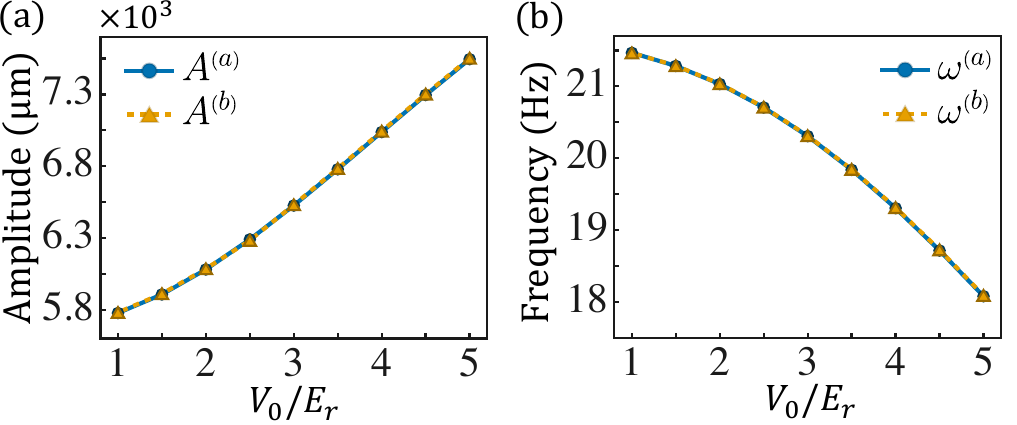}
\par\end{centering}
\caption{Comparison of the amplitudes  and frequencies between $R_{x}^{{\rm s}\ua}(t) = A^{\ua} [1- \cos\omega^\ua t]$ and $R_{x}^{{\rm c}\prts{b}}(t)=A^{\prts{b}}[1-\cos\omega^\prts{b} t]$ for various lattice depths.}
\label{fig:GPampfreq}
\end{figure}

\begin{figure}[t]
\begin{centering}
\includegraphics[width=8.6cm]{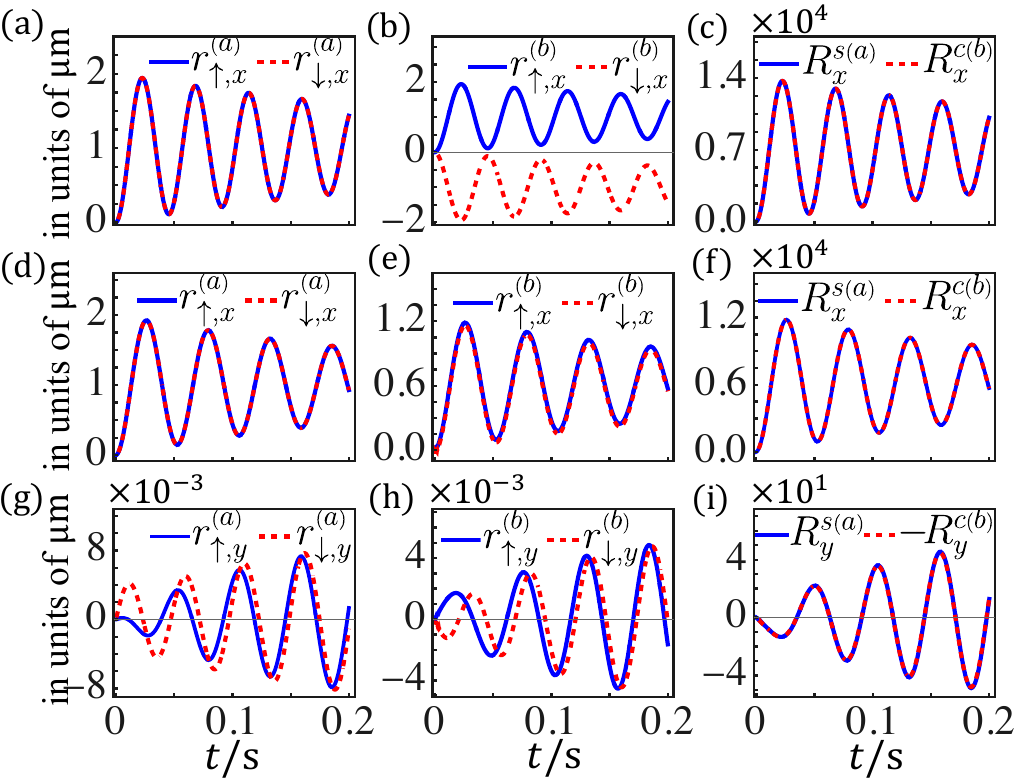}
\par\end{centering}
\caption{The dissipative quench dynamics are simulated by the dissipative GP equation, where the dissipation parameter is taken to be $\gamma=0.005$; The rest of the  parameters are the same as those in Fig.~\ref{fig:GP1}. }
\label{fig:GP2}
\end{figure}

This means that the longitudinal spin displacement resulting from quench protocol (a) must be exactly the same as the longitudinal charge displacement resulting from quench protocol (b), while the corresponding transverse spin and charge displacements must be exactly the opposite. 
To appreciate the nontrivial nature of the relations in Eq.~(\ref{Rrelations}), we first observe that the charge and spin displacements in experiments can be simply read out from the center of mass (c.m.) of the spin components $r_{\uparrow,\mu}$ and $r_{\downarrow,\mu}$,  viz., 
\begin{align}
\label{Rc}
R^{{\rm s}\ua}_{\nu}(t)& = {N_\uparrow}r^{\ua}_{\uparrow,\nu}(t) - {N_\downarrow}r^{\ua}_{\downarrow,\nu}(t); \\
R^{{\rm c}\ub}_{\mu}(t) &= {N_\uparrow}r^{\ub}_{\uparrow,\mu}(t) + {N_\downarrow}r^{\ub}_{\downarrow,\mu}(t),
\label{Rsc}
\end{align}
where $N_\uparrow$ and $N_\downarrow$ are the numbers of spin-up and spin-down atoms respectively. The first relation in Eq.~(\ref{Rrelations}) can be understood if there is no atomic interaction and no SOC at all, in which case the dynamics of the c.m. of the two spin components are completely independent. It is clear from the two protocols that $r^{\ua}_{\uparrow,x}(t) = r^{\ub}_{\uparrow,x}(t)$ and $r^{\ua}_{\downarrow,x}(t) = - r^{\ub}_{\downarrow,x}(t)$, which yields $R^{{\rm s}\ua}_{x}(t) =  R^{{\rm c}\ub}_{x}(t)$ in light of Eqs.~(\ref{Rc})-(\ref{Rsc}) [see Fig.~\ref{fig:GP1}(a)-(c)].  Now, the generalized Onsager relations predicts that this must always be the case for any system parameter as long as the ground state of the system has the required symmetry. To verify this, we numerically simulate the c.m. dynamics in these two quench protocols using the time-dependent Gross-Pitaevskii equation for spinor condensates~\cite{Kawaguchi2012}
\begin{align}
(i-\gamma){\pa_t \Phi_\sigma}=\sum_{\sigma'} [( h_{\sigma\sigma'}+   {\tilde V}_{\rm tr}\delta_{\sigma\sigma'})\Phi_{\sigma'}+g_{\sigma\sigma'}|\Phi_{\sigma'}|^2\Phi_\sigma ],
\end{align}
where $\gamma$ is a phenomenological parameter that is often used to account for the effect of thermal dissipation and $\bPhi(\br,t) = (\Phi_\uparrow(\br,t),\, \Phi_\downarrow(\br,t))^T$ is the time-dependent condensate wave function. We first consider the case of zero temperature corresponding to $\gamma = 0$. The c.m. dynamics of the spin components are calculated by 
\begin{align}
r_{\sigma,\mu} (t)= \frac{1}{N_\sigma}\int d\br r_\mu |\Phi_{\sigma}(\br,t)|^2
\end{align} and are shown in the first [protocol (a)] and the second column [protocol (b)] of Fig.~\ref{fig:GP1}. In contrast to independent two-component case [Fig.~\ref{fig:GP1}(a)-(b)], we see that $r^{\ua}_{\uparrow,x}(t) \neq  r^{\ub}_{\uparrow,x}(t)$ and $r^{\ua}_{\downarrow,x}(t) \neq - r^{\ub}_{\downarrow,x}(t)$ when atomic interactions and SOC are present [Fig.~\ref{fig:GP1}(d)-(e)]. In fact, the strong coupling of the two spin components is most clearly seen in Fig.~\ref{fig:GP1}(e), where the c.m. dynamics of the two components are almost synchronized even though the spin force is meant to drive them apart. Nevertheless, the first relation in Eq.~(\ref{Rrelations}) is perfectly satisfied as shown in Fig.~\ref{fig:GP1}(f). Furthermore, we find that $R_{x}^{{\rm s}\ua}(t)$ and $R_{x}^{{\rm c}\ub}(t)$ exhibit undamped behavior described by 
\begin{align}
R_{x}^{{\rm s}\ua}(t) &= A^{\ua} [1- \cos\omega^\ua t]; \\ R_{x}^{{\rm c}\prts{b}}(t)&=A^{\prts{b}}[1-\cos\omega^\prts{b} t].
\end{align}
By comparing the amplitudes and frequencies of $R_{x}^{{\rm s}\ua}(t)$ and $R_{x}^{{\rm c}\ub}(t)$ for various lattice depths (see Fig.~\ref{fig:GPampfreq}), we have shown that the validity of the generalized Onsager relation is indeed independent of specific system parameters. The second relation in Eq.~(\ref{Rrelations}) describing the transverse charge-spin dynamics is similarly confirmed in Fig.~\ref{fig:GP1}(g)-(i), where we find that $R^{{\rm s}\ua}_{ y} (t)$ and $ R^{{\rm c}\ub}_{y} (t)$ are exactly opposite to each other.  In this respect,  protocol (a) and (b) actually realize the spin Hall effect and its reciprocal process, the inverse spin Hall effect, respectively. 

Finally, we have verified that Eq.~(\ref{Rrelations}) holds even when thermal dissipation is present, suggesting that these relations can be experimentally tested at finite temperatures. As demonstrated in Fig.~\ref{fig:GP2}, the presence of dissipation leads to appreciable damping of the longitudinal oscillations. It also reduces the amplitude of the transverse dynamics as can be seen by a comparison of the third row of Fig.~\ref{fig:GP2} with that of Fig.~\ref{fig:GP1}. Despite the significant effects of the dissipation, the generalized Onsager reciprocal relations are perfectly obeyed, as can be seen in (c), (f), and (i) of Fig.~\ref{fig:GP2}. 

\section{Conclusions}
\label{conclusion}
In conclusion, we have shown that the Onsager reciprocal relations of charge and spin transport can be generalized to systems that break the time-reversal symmetry $\calT$ but preserve certain composite $\calO\calT$ symmetries. As a conceptual example, we have established these relations for various ground states of a simple spin-orbit-coupled model in quantum gases, by identifying the relevant $\calO$ symmetry operations. We have further calculated the charge-spin transport coefficients and verified these relations explicitly. In addition, we have discussed experimental implications of these relations in the context of quench dynamics frequently performed in quantum gas experiments. For many solid state materials, our generalization reveals a hidden symmetry or antisymmetry between the charge-spin cross-coupling coefficients that could be useful in future studies of spintronics. Lastly, our proof of the generalized Onsager relations of charge and spin transport can be modified to discuss other cross-coupling transport phenomena in time-reversal symmetry breaking systems once appropriate $\calO\calT$ symmetries are identified.

\section*{Acknowledgements}  This work is supported by National Key R$\&$D Program of China (Grant No. 2022YFA1404103), Natural Science Foundation of China (Grant No.~12474264, No.~12174455), Guangdong Provincial Quantum Science Strategic Initiative (Grant No.~GDZX2404007), Natural Science Foundation of Guangdong
Province (Grant No. 2021B1515020026), and Guangdong
Basic and Applied Basic Research Foundation (Grant No.
2023B1515040023). S.Z. acknowledges support from HK GRF (Grant No. 17306024), CRF (Grants No. C6009-20G, No. C7012-21G, No. C4050-23GF), and a RGC Fellowship Award No.
HKU RFS2223-7S03. 

\appendix
\setcounter{section}{0}  
\renewcommand{\theequation}{A\arabic{equation}}
\setcounter{equation}{0}  
\section{Onsager-Casimir relations}
\label{AppA}
In this appendix, we discuss the conceptual difference between our generalized Onsager relations and the well-known Onsager-Casimir relations~\cite{Casimir};   the latter is a generalization of the Onsager relations to electronic systems in which the time-reversal symmetry is explicitly broken by the presence of an external magnetic field $\boldsymbol B$.  For concreteness, we restrict ourselves to the charge and spin cross-coupling coefficients with respect to the ground state of an electronic system.   The Hamiltonian for the system under the magnetic field is 
\begin{align}
\hat H(\bB) = &\hat H_0  -\frac{e}{2m}\sum_{j}(\hat \bp_j\cdot \bA + \bA\cdot \hat\bp_j ) \nn \\
 &- g\mu_{\rm B} \bB \cdot  \sum_j \hat \bs_j,
\label{HB}
\end{align}
where $\hat H_0$ is the Hamiltonian of the system without the magnetic field (assumed to be time-reversal invariant itself), $\bA$ is the vector potential, $\mu_{\rm B}$ is the Bohr magneton and $g$ is the Land{\'e} g-factor. If the system has no spontaneous symmetry-breaking in the absence of the magnetic field, the ground state $|\Phi(\bB)\ra$ can be viewed as an analytic function of the magnetic field $\bB$ which can be continuously varied across zero. In this case, the Onsager-Casimir relations of the charge and spin transport read
\begin{align}
 \sigma^{\rm sc}_{\mu\nu}  (\boldsymbol B) = - \sigma^{\rm cs}_{\nu\mu}  (-\boldsymbol B),
\label{OC}
\end{align}
where 
\begin{align}
\sigma^{\rm sc}_{\mu\nu} ( \boldsymbol B)  &= i\! \int_{0}^\infty \! \!  d t e^{i\omega t}\la\Phi(\bB)|     [  {\hat J}^{\rm s}_{\mu} (t) , \hat R^{\rm c}_{\nu} (0)  ]  |\Phi(\bB)\ra; \nn  \\ 
\sigma^{\rm cs}_{\nu\mu} (-\boldsymbol B) &= i \! \int_{0}^\infty \! \! d t e^{i\omega t} \la \Phi(-\bB)|   [  {\hat J}^{\rm c}_{\nu} (t) , \hat R^{\rm s}_{\mu} (0)  ]  | \Phi(-\bB)\ra. \nn 
\end{align}
Naturally, the Heisenberg operators are also functions of $\bB$ due to their dependence on $\hat H(\bB)$, e.g. $\hat J^{\rm s}_\mu(t) = e^{i\hat H(\bB) t}\hat J^{\rm s}_\mu(0) e^{-i\hat H(\bB) t}$. In the limit of $\bB\rightarrow 0$, the Onsager-Casimir relations reduce to the Onsager relations of Eq.~(\ref{onsager}) in the main text. In this sense, we see that the Onsager-Casimir relations contain the Onsager relations as a special case. The Onsager-Casimir relation can be derived from the following more general relation which is valid for an arbitrary Hamiltonian $\hat H$ and regardless of whether the system has spontaneous symmetry-breaking
\begin{align}
\left. \sigma^{\rm sc}_{\mu\nu}\right|_{\hat H, \Phi}  = - \left. \sigma^{\rm cs}_{\nu\mu}\right |_{\hat{\tilde H}, \tilde \Phi},
\label{goc}
\end{align}
where the left-hand side is evaluated under $\hat H$ and $|\Phi\ra$ and the right-hand side is evaluated under the time-reversal-symmetry-transformed Hamiltonian and state
\begin{align}
\hat {\tilde H} = \mathcal T \hat H \mathcal T^{-1}; \qquad|\tilde \Phi\ra = \calT | \Phi\ra.
\end{align}
More specifically, we have
\begin{align}
\label{ocsigmakubo1g}
\left. \sigma^{\rm sc}_{\mu\nu} \right|_{\hat H, \Phi}  &\equiv i\int_{0}^\infty d t \la\Phi|     [  {\hat J}^{\rm s}_{\mu} (t) , \hat R^{\rm c}_{\nu} (0)  ]  |\Phi\ra e^{i\omega t} ;  \\
 \left. \sigma^{\rm cs}_{\nu\mu}\right |_{\hat{\tilde H}, \tilde \Phi} &\equiv i \int_{0}^\infty d t \la \tilde\Phi|   [  \hat { {\tilde J}}^{\rm c}_{\nu} (t) , \hat {\tilde R}^{\rm s}_{\mu} (0)  ]  | \tilde \Phi\ra e^{i\omega t},
\label{ocsigmakubo2g}
\end{align}
where $\hat { {\tilde J}}^{\rm c}_{\nu} (t)  = e^{i\hat {\tilde H} t}\hat { { J}}^{\rm c}_{\nu} (0)e^{-i\hat {\tilde H} t}$. The relation in (\ref{goc}) can be derived straightforwardly by applying the time-reversal symmetry operation $\mathcal T$ to both the state and the Heisenberg operators in (\ref{ocsigmakubo1g}) and then following similar steps as those from Eq.~(\ref{ROT}) to Eq.~(\ref{gonsager}) in the main text. To obtain the Onsager-Casimir relation from (\ref{goc}) we take $\hat H(\bB)$ in (\ref{HB}) as $\hat H$ and find 
\begin{align}
\hat {\tilde H} = \mathcal T \hat H (\bB) {\mathcal T}^{-1} = \hat H (-\bB).
\label{tildeH}
\end{align}
 Furthermore it can be shown that $\mathcal T|\Phi(\bB)\ra$ is an eigenstate of $\hat H(-\bB)$, i.e., 
\begin{align}
|\tilde \Phi\ra \equiv \mathcal T | \Phi(\bB)\ra = |\Phi(-\bB)\ra.
\label{tildePhi}
\end{align}
Note that the above relation holds only if the system has no time-reversal symmetry breaking in the absence of the magnetic field. This can be seen from the fact that  the above equation demands that the state $ |\Phi(0)\ra$ be time-reversal invariant in the limit of $\bB \rightarrow 0$. Using (\ref{tildeH}) and (\ref{tildePhi}) in (\ref{goc})  we then arrive at the Onsager-Casimir relation in (\ref{OC}). 

In the special case of $\bB = 0$, we immediately have $\hat {\tilde H} = \hat H$ and $|\tilde \Phi\ra = |\Phi\ra$, and the relation in (\ref{goc}) becomes the Onsager relation. From the perspective of (\ref{goc}) we see a conceptual difference between the Onsager relations and the Onsager-Casimir relations; the two cross-coupling coefficients in the former are determined with respect to the same Hamiltonian and state while those in the latter are determined with respect to different Hamiltonians and states. This critical difference remains between our generalized Onsager relations and the Onsager-Casimir relations. Using the notation of (\ref{goc}), our generalized Onsager relations can be written as
\begin{align}
\left. \sigma^{\rm sc}_{\mu\nu}\right|_{\hat H, \Phi}  = (-1)^{I_{\mu\nu}} \left. \sigma^{\rm cs}_{\nu\mu}\right |_{\hat{ H},  \Phi}.
\label{gor}
\end{align}
Thus,  just like the Onsager relations they connect two cross-coupling coefficients of the same Hamiltonian and state.

Besides the conceptual difference, our generalized Onsager relations and the Onsager-Casimir relations have different domains of applicability. We have shown that the Onsager-Casimir relation in (\ref{OC}) only applies to the case where the time-reversal symmetry is explicitly broken by the presence of an external magnetic field $\boldsymbol B$. However, there are many other scenarios where the time-reversal symmetry is broken not by an external magnetic field.  Classic examples are magnets (ferromagnetic, antiferromagnetic and altermagnetic) in which the time-reversal symmetry $\mathcal T$ is spontaneously broken by the formation of magnetic orders. Other examples include certain cold atomic systems where $\mathcal T$ is broken by various types of synthetic gauge fields. The main purpose of our work is to point out that the known Onsager reciprocal relations of charge and spin transport do not apply to these systems but generalized Onsager relations can be established if the system preserves certain combined $\mathcal O\mathcal T$ symmetry.   

Finally, we provide an alternative derivation of our generalized Onsager relations which elucidates the role of the additional $\mathcal O$ operation. We begin with the relation in (\ref{goc}) which remains valid. Since time-reversal symmetry is broken, we find that $|\tilde \Phi\ra = \mathcal T |\Phi\ra \neq |\Phi\ra$. In order to relate the right-hand side of (\ref{goc}) back to the cross-coupling coefficient evaluated under the same $\hat H$ and $|\Phi\ra$, we may perform an additional $\mathcal O$ operation on the state and Heisenberg operators in (\ref{ocsigmakubo2g}).  Making use of the $\mathcal O\mathcal T $ symmetry of the Hamiltonian and the state, i.e., $(\mathcal O\mathcal T) \hat H (\mathcal O\mathcal T)^{-1} = \hat H$ and $(\mathcal O\mathcal T)|\Phi\ra = e^{i\theta} |\Phi\ra$, as well as the transformations in Eq.~(\ref{O}) of the main text, we find that
\begin{align}
 \left. \sigma^{\rm cs}_{\nu\mu}\right |_{\hat{\tilde H},\tilde \Phi}  =  (-1)^{I_{\mu\nu}-1} \left. \sigma^{\rm cs}_{\nu\mu}\right |_{\hat H, \Phi}.
\end{align}
Substituting the above equation into (\ref{goc}) we then arrive at the generalized Onsager relations in (\ref{gor}). We thus see that the role of the additional $\mathcal O$ operation is to bring the time-reversed state back to the original state so that an relationship between cross-coupling coefficients of the same state can be established.

\renewcommand{\theequation}{B\arabic{equation}}
\setcounter{equation}{0}  
\section{Derivation of alternative Kubo formulas}
\label{AppB}
In this appendix, we derive alternative Kubo formulas of charge-spin cross-coupling transport tensors in a lattice system, i.e., Eqs.~(\ref{sigmakubo3})-(\ref{sigmakubo4}) of the main text, and show that they are equivalent to Eqs.~(\ref{sigmakubo1})-(\ref{sigmakubo2}) of the main text.  Before doing so, we first show that Eqs.\,(\ref{sigmakubo1})-(\ref{sigmakubo2}) of the main text can also be expressed in terms of the more familiar current-current correlation functions.
We begin with the charge-spin response coefficient $\sigma_{\nu\mu}^{\text{cs}}(\omega)$ in Eq.~(\ref{sigmakubo2}) of the main text, which can be written as
\begin{align}
\sigma^{\text{cs}}_{\nu\mu}(\omega)  = -\chi_{J^{\rm c}_\nu,R^{\rm s}_\mu}(\omega),
\label{scschi}
\end{align}
where  $\chi_{J^{\rm c}_\nu,R^{\rm s}_\mu}(\omega)$ is the Fourier transform of 
\begin{align}
\chi_{J^{\rm c}_\nu,R^{\rm s}_\mu}(t-t')=-i\theta(t-t')\langle[\hat{J}^{\rm c}_\nu(t),\hat{R}^{\rm s}_\mu(t')]\rangle.
\end{align}
Taking the  derivative of $\chi_{J^{\rm c}_\nu,R^{\rm s}_\mu}(t-t')$ with respect to $t'$ we obtain
\begin{equation}
    \frac{d}{dt'}\chi_{J^{\rm c}_\nu,R^{\rm s}_\mu}(t-t')= i\delta(t-t')\langle[\hat J^{\rm c}_\nu,\,\hat R^{\rm s}_\mu]\rangle + \chi_{J^{\rm c}_\nu,J^{\rm s}_\mu}(t-t'), \label{eq:XjsRc1}
\end{equation}
where $\hat J^{\rm s}_{\mu} = \frac{d}{dt}\hat R^{\rm s}_\mu$ has been used and 
\begin{align}
\chi_{J^{\rm c}_\nu, J^{\rm s}_\mu}(t-t')=-i\theta(t-t')\langle[\hat{J}^{\rm c}_\nu(t),\hat{J}^{\rm s}_\mu(t')]\rangle.
\end{align}
Performing the Fourier transform, \eqref{eq:XjsRc1} becomes
\begin{equation}
    i\omega\chi_{J^{\rm c}_\nu,R^{\rm s}_\mu}(\omega)= i\langle[\hat J^{\rm c}_\nu,\,\hat R^{\rm s}_\mu]\rangle + \chi_{J^{\rm c}_\nu,J^{\rm s}_\mu}(\omega). \label{eq:XjsRc2}
\end{equation}
Substituting \eqref{eq:XjsRc2} into \eqref{scschi}, we obtain
\begin{equation}
    \sigma^{\text{cs}}_{\nu\mu}(\omega) = \frac{i}{\omega}\bigg[i\la [\hat J^{\rm c}_\nu,\,\hat R^{\rm s}_\mu] \ra  + \chi_{J^{\rm c}_\nu,J^{\rm s}_\mu}(\omega) \bigg]. \label{eq:XjsRc3}
\end{equation}
By analogy to the charge transport, the first term of \eqref{eq:XjsRc3} corresponds to the contribution of a diamagnetic current, while the second term given by the charge current-spin current correlation function  corresponds to the contribution of a paramagnetic current. Similarly, the spin-charge  transport coefficient can also be expressed in terms of current-current correlation function as 
\begin{equation}
    \sigma^{\text{sc}}_{\mu\nu}(\omega) = \frac{i}{\omega}\bigg[i\langle[\hat J^{\rm s}_\mu,\,\hat R^{\rm c}_\nu]\rangle + \chi_{J^{\rm s}_\mu,J^{\rm c}_\nu}(\omega) \bigg]. \label{eq:XjsRc4}
\end{equation}
However, both Eqs.\,\eqref{eq:XjsRc3}-\eqref{eq:XjsRc4} and Eqs.\,(\ref{sigmakubo1})-(\ref{sigmakubo2}) of main text are inconvenient to use for a lattice system with periodic boundary conditions.

Now we proceed to demonstrate the equivalence between Eqs.\,(\ref{sigmakubo1})-(\ref{sigmakubo2}) and (\ref{sigmakubo3})-(\ref{sigmakubo4}) of main text. We first consider the charge-spin response coefficients $\sigma_{\mu\nu}^{\text{sc}}(\omega)$. In the presence of an external potential $V^{\rm c}(\br,t)$ that couples to the charge, the spin current density is given by linear response theory as
\begin{align}
    j^{\rm s}_{\mu}(\bm q,\omega) = \chi_{j^{\rm s}_{\mu},\rho^{\rm c}}(\bq,\omega)V^{\rm c}(\bq,\omega), \label{eq:jwq1}
\end{align}
where $V^{\rm c}(\bq,\omega)$ is the Fourier transform of $V^{\rm c}(\br,t)$, 
$\chi_{A,B}(\bm q,\omega)$ is the Fourier transform of the retarded response function
\begin{align}
\chi_{A,B}(\bm q,t-t')\equiv -i\theta(t-t')\langle[\hat A(\bm q,t),\hat B(-\bm q,t')]\rangle,
\end{align}
and $A(\bq,\omega)\equiv \int dt\la \hat A(\bq,t)\ra e^{i\omega t}$ as per standard notation. 
Here $\hat j^{\rm s}_\mu(\bm q,t)$ and $\hat \rho^{\rm c}(\bm q,t)$ are the Fourier transforms of spin current operator
\begin{equation}
    \hat j_{\mu}^{\rm s}(\bm r,t) = \frac{d}{dt}\big[\hat\bpsi^\dagger(\bm r,t)r_\mu s_z\hat\bpsi(\bm r,t)\big]
\end{equation}
and density operator 
\begin{equation}
    \hat \rho^{\rm c}(\bm r,t)=\hat\bpsi^\dagger(\bm r,t)\hat\bpsi(\bm r,t)
\end{equation}
respectively, where $\hat\bpsi(\bm r,t)=[\hat\psi_\uparrow(\bm r,t),\hat\psi_\downarrow(\bm r,t)]^T$. Making use of 
\begin{align}
    \hat j_{\mu}^{\rm s}(\bm q,t) &= \frac{d}{dt}\int d\bm r\, \hat\psi^\dagger(\bm r,t)r_\mu s_z\hat\psi(\bm r,t) e^{-i\bm q\cdot\bm r} \nn \\ &= \frac{d}{dt}i\partial_{q_{\mu}}\hat{\rho}^{\rm s}(\bm{q},t) \label{eq:jq1}
\end{align}
where
\begin{equation}
    \hat{\rho}^{\rm s}(\bm{q},t) = \int d\bm r\, \hat\bpsi^\dagger(\bm r,t) s_z\hat\bpsi(\bm r,t) e^{-i\bm q\cdot\bm r} \label{eq:rhosq}
\end{equation} 
is the Fourier transform of the spin density, we can alternatively write
\begin{align}
    j_{\mu}^{\rm s}(\bm q,\omega)&= \omega  \partial_{q_\mu} \rho^{\rm s}(\bm q,\omega) \nn \\ &= \omega  \partial_{q_\mu}[\chi_{\rho^{\rm s},\rho^{\rm c}}(\bm q,\omega) V^{\rm c}(\bm q,\omega)]\label{eq:jq2}
\end{align}
where we used the linear response result
\begin{equation}
    \rho^{\rm s}(\bm q,\omega) = \chi_{\rho^{\rm s},\rho^{\rm c}}(\bm q,\omega) V^{\rm c}(\bm q,\omega).
\label{rhosXV}
\end{equation}
We now have two equivalent expressions, (\ref{eq:jwq1}) and (\ref{eq:jq2}), for the spin current density. We shall show that in the limit of $\bq\rightarrow 0$ the first leads to Eq.~(\ref{sigmakubo1}) and the second to Eq.~(\ref{sigmakubo3}) of the main text. To do so, we first note that the external potential $V^{\rm c}(\bm r, t)$ generates a force $\bm F^{\rm c}(\bm r, t)=-\nabla V^{\rm c}(\bm r, t)$ from which we find 
\begin{align}
    \bF^{\rm c}(\bq,\omega) = -i\bq V^{\rm c}(\bq,\omega)
    \label{Fcfourier}
\end{align}
and
\begin{equation}
 V^{\rm c}(\bm q,\omega)=i\sum_\nu\frac{q_\nu}{q^2}F^{\rm c}_\nu(\bm q,\omega). \label{eq:VbyF}
\end{equation}
Substituting \eqref{eq:VbyF} into \eqref{eq:jwq1}, we obtain
\begin{gather}
    j^{\rm s}_{\mu}(\bm q,\omega) = \sum_\nu i\frac{q_\nu}{q^2}\chi_{j^{\rm s}_{\mu},\rho^{\rm c}}(\bq,\omega)F^{\rm c}_\nu(\bm q,\omega). \label{jsqw1}
    \end{gather}
From the definition of the transport tensor 
\begin{align}
 j^{\rm s}_{\mu}(\bm q,\omega) = \sum_\nu  \sigma_{\mu\nu}^{\text{sc}}(\bq,\omega)F^{\rm c}_\nu(\bm q,\omega),
 \end{align}
we find that
    \begin{align}
 \sigma_{\mu\nu}^{\text{sc}}(\bm q,\omega)=i\frac{q_\nu}{q^2}\chi_{j^{\rm s}_{\mu},\rho^{\rm c}}(\bq,\omega)  \label{sigma_scqw}.
\end{align}
Since $\hat \rho^{\rm c}(\bm q=0,0)$ is the total particle number operator and commutes with the total current $\hat j^{\rm s}_\mu(\bm q =0,t)$, i.e., $[\hat j_\mu^{\rm s}(0,t),\hat\rho^{\rm c}(0,0)]=0$,  we find that $\chi_{j^{\rm s}_{\mu},\rho^{\rm c}}(0,\omega)=0$. Thus we have
\begin{align}
    \sigma_{\mu\nu}^{\text{sc}}(\omega) \equiv &\lim_{\bq\rightarrow 0 }\sigma_{\mu\nu}^{\text{sc}}(\bq,\omega) \nn \\
   =&i\lim_{q_\nu\to0}\lim_{q_{\nu_\perp}\to0}\frac{q_\nu}{q^2}[\chi_{j^{\rm s}_{\mu},\rho^{\rm c}}(\bq,\omega) - \chi_{j^{\rm s}_{\mu},\rho^{\rm c}}(0,\omega)] \nn \\ 
   = &i\partial_{q_\nu}\chi_{j^{\rm s}_{\mu},\rho^{\rm c}}(\bm q,\omega)\big|_{\bm q=0} \nn \\
   =&\int_0^\infty  dt \la [\hat j^{\rm s}_\mu(\bm q,t),\partial_{q_\nu}\hat\rho^{\rm c}(-\bq, 0)]\ra e^{i\omega t}\Big|_{\bm q=0} \nn \\ +& \int_0^\infty  dt \la [\partial_{q_\nu}\hat j^{\rm s}_\mu(\bm q,t), \hat\rho^{\rm c}(-\bq, 0)]\ra e^{i\omega t}\Big|_{\bm q=0}, \label{sigma_sc1}
\end{align}
where $\nu_\perp$ denotes the direction orthogonal to $\nu$ in the $xy$-plane such that $q^2 = q_\nu^2+q_{\nu_\perp}^2$. We note that the second term of \eqref{sigma_sc1} represents a response induced by a constant external potential which must vanish; the remaining term is precisely the right hand side of Eq.~(\ref{sigmakubo1}) in the main text, expressed in second quantization form.
Next we can write \eqref{eq:jq2} as
\begin{align}
    j_{\mu}^{\rm s}(\bm{q},\omega)
    =& \omega\left[\partial_{q_{\mu}}\chi_{\rho^{\rm s},\rho^{\rm c}}(\bm{q},\omega)\right]V^{\rm c}(\bm{q},\omega) \nn \\
    &+\omega\chi_{\rho^{\rm s},\rho^{\rm c}}(\bm{q},\omega)\partial_{q_{\mu}}V^{\rm c}(\bm{q},\omega).\label{eq:jq3}
\end{align}
It can be shown that $\chi_{\rho^{\rm s},\rho^{\rm c}}( 0,\omega)=0$ due to the fact that $[\hat\rho^{\rm s}(0,t),\hat \rho^{\rm c}(0,0)]=0$. Thus in the limit that $\bq \rightarrow 0$ 
we only need to consider the first term of \eqref{eq:jq3}. We shall show that 
\begin{equation}
    \chi_{\rho^{\rm s},\rho^{\rm c}}(\bm{q},\omega)=\frac{1}{\omega}\sum_{\nu}q_{\nu}\chi_{\rho^{\rm s},j^{\rm c}_{\nu}}(\bm{q},\omega),\label{eq:Xszrho2}
\end{equation}
where
$\hat j_{\nu}^{\rm c}(\bm q,t) = \frac{d}{dt}\int d\bm r\, \hat\bpsi^\dagger(\bm r,t)r_\nu \hat\bpsi(\bm r,t) e^{-i\bm q\cdot\bm r}$
is the Fourier transform of charge current density operator. 
Taking the derivative of $\chi_{\rho^{\rm s},\rho^{\rm c}}(\bq,t-t')$ with respect to $t'$, we have
\begin{align}
    \frac{d}{dt'}\chi_{\rho^{\rm s},\rho^{\rm c}}(\bm{q},t-t')&= i\delta(t-t')\langle[\hat\rho^{\rm s}(\bm{q},t),\,\hat\rho^{\rm c}(-\bm{q},t)]\rangle \nn \\
    -i\theta(t-t')&\langle[\hat\rho^{\rm s}(\bm{q},t),\,\partial_t\hat \rho^{\rm c}(-\bm{q},t')]\rangle. \label{eq:Xszrho0}
\end{align}
The commutator in the first term vanishes. Using the continuity equation 
$
 \partial_t\hat\rho^{\rm c}(\bm q,t)+i\bm q\cdot \hat{\bm j}^{\rm c}(\bm q,t)=0,  
$
\eqref{eq:Xszrho0} becomes 
\begin{align}
    \frac{d}{dt'}\chi_{\rho^{\rm s},\rho^{\rm c}}(\bm{q},t-t')
    =&\sum_{\nu}iq_{\nu}\chi_{\rho^{\rm s},j^{\rm c}_{\nu}}(\bm{q},t-t')\nonumber \\
    = & \sum_{\nu}iq_{\nu}\int\chi_{\rho^{\rm s},j^{\rm c}_{\nu}}(\bm{q},\omega)e^{-i\omega(t-t')}d\omega. \label{eq:Xszrho1}
\end{align}
On the other hand, we also have 
\begin{align}
\frac{d}{dt'}\chi_{\rho^{\rm s},\rho^{\rm c}}(\bm{q},t-t')=\int i\omega\chi_{\rho^{\rm s},\rho^{\rm c}}(\bm{q},\omega)e^{-i\omega(t-t')}d\omega. 
\end{align} 
Comparing this with \eqref{eq:Xszrho1}, we arrive at (\ref{eq:Xszrho2}). Substituting \eqref{eq:Xszrho2} into \eqref{eq:jq2}, we find 
\begin{align}
    j^{\rm s}_{\mu}(\bm{q}\to 0,\omega)= &\lim_{\bq\rightarrow 0}\sum_{\nu}[\delta_{\mu\nu}\chi_{\rho^{\rm s},j^{\rm c}_{\nu}}(\bm{q},\omega)\nn \\
    &+q_{\nu}\partial_{q_{\mu}}\chi_{\rho^{\rm s},j^{\rm c}_{\nu}}(\bm{q},\omega)]V^{\rm c}(\bm{q},\omega).
\end{align}
We note that the first term vanishes because $\chi_{j_{\nu}^{\rm c},\rho^{\rm{s}}}(0,\omega)=0$. This latter expression vanishes because it represents the charge current response induced by a uniform Zeeman field. 
Lastly, using (\ref{Fcfourier}), we find 
\begin{equation}
    j^{\rm s}_{\mu}(\bq\rightarrow 0,\omega) = \lim_{\bm{q}\to 0}\sum_{\nu}i\partial_{q_{\mu}}\chi_{\rho^{\rm s},j^{\rm c}_{\nu}}(\bm{q},\omega)F^{\rm c}_\nu(\bm{q},\omega)
\end{equation}
which implies that 
\begin{equation}
\sigma_{\mu\nu}^{\rm sc}(\omega)=i\partial_{q_{\mu}}\chi_{\rho^{\rm s},j^{\rm c}_{\nu}}(\bm{q},\omega)\Big|_{\bm{q}=0}.\label{eq:Sigma_sc}
\end{equation}
This is nothing but Eq.~(\ref{sigmakubo3}) of the main text. 

Now, we consider the charge-spin response coefficients $\sigma_{\nu\mu}^{\rm{cs}}(\omega)$. In the presence of a potential $V^{\rm s}(\br,t)$ that couples to the spin, the perturbation to the Hamiltonian is $s_zV^{\rm s}(\br,t)$. 
The charge current induced by such a spin-dependent potential is given by linear response theory as
\begin{equation}
j_{\nu}^{\rm c}(\bm{q},\omega)=\chi_{j_{\nu}^{\rm c},\rho^{\rm s}}(\bm{q},\omega)V^{\rm s}(\bm{q},\omega).
\label{jcnu}
\end{equation}
Just like the case of the charge force, we have  
$ V^{\rm s}(\bm q)=\sum_\mu iF_{\mu}^{\rm s}(\bm q,\omega)q_\mu/q^2$. Using this in (\ref{jcnu}) we find 
\begin{equation}
j_{\nu}^c(\bm{q},\omega)=\sum_{\mu}\chi_{j_{\nu}^{\rm c},\rho^{\rm s}}(\bm{q},\omega)\frac{iq_{\mu}}{q^{2}}F_{\mu}^{\rm s}(\bm{q},\omega).\label{eq:Xjusz1}
\end{equation}
This implies that
\begin{align}
    \sigma_{\nu\mu}^{\rm{cs}}(\omega)&=\lim_{\bm q\to 0}\frac{iq_{\mu}\chi_{j_\nu^{\rm c},\rho^{\rm s}}(\bm{q},\omega)}{q^{2}} \nn \\
    &=i\partial_{q_{\mu}}\chi_{j_{\nu}^{\rm c},\rho^{\rm s}}(\bm q,\omega)\Big|_{\bm q=0},\label{eq:Sigma_cs}
\end{align}
where $\chi_{j_{\nu}^{\rm c},\rho^{\rm s}}(0,\omega)=0$ is used. We see that this is simply Eq.~(\ref{sigmakubo4}) of the main text. Now, \eqref{eq:Sigma_cs} can be written as
\begin{align}
    \sigma_{\nu\mu}^{\rm{cs}}(\omega) = & \partial_{q_\mu}\Big[\int_0^\infty dt\langle[\hat j_\nu^{\rm c}(\bq,t),\hat\rho^{\rm s}(-\bq,0)]\rangle e^{i\omega t}\Big]_{\bm q=0} \notag \\
    = & \int_0^\infty dt\langle[ \partial_{q_\mu}\hat j_\nu^{\rm c}(\bq,t),\hat\rho^{\rm s}(-\bq,0)]\rangle e^{i\omega t}\Big|_{\bm q=0} \nn \\
    +& \int_0^\infty dt\langle[\hat j_\nu^{\rm c}(\bq,t),\partial_{q_\mu}\hat\rho^{\rm s}(-\bq,0)]\rangle e^{i\omega t}\Big|_{\bm q=0}.
    \label{sigmaw2}
\end{align}
The first term of \eqref{sigmaw2} again represents a current-related response induced by a uniform Zeeman field and so should vanish; the remaining term is  exactly Eq.~(\ref{sigmakubo2}) of main text in second quantization.

Both \eqref{eq:Sigma_sc} and \eqref{eq:Sigma_cs} are convenient to use in a lattice system. The derivatives of the response functions can be carried out numerically by the method of central difference, or analytically by expressing the response functions in terms of the Green's functions. 

\renewcommand{\theequation}{C\arabic{equation}}
\setcounter{equation}{0}  
\section{Calculation of the charge-spin transport coefficients for the Fermi gas }
\label{tcfermi}
In this appendix, we provide more details on the band structure of the model~\cite{Huang2021} and the evaluation of the charge-spin cross-coupling transport coefficients for a noninteracting Fermi gas. We first determine the Bloch bands of the noninteracting Hamiltonian using the plane-wave expansion.  For this purpose, $\bm a_1 = (1,1)\pi/k_L$ and $\bm a_2 = (-1,1)\pi/k_L$ are chosen as the primitive vectors, and the reciprocal primitive vectors vectors $\bm b_i$ are determined by $\bm a_i\cdot\bm b_i = 2\pi\delta_{ij}$. We denote the plane-wave basis by $\ket{\bm k,\!\bm Q,\sigma}=\ket{\bm k,\!\bm Q}\!\otimes\!\ket{\sigma}$, where $\bm k\in \rm{BZ}$, $ \bm Q = m\bm b_1+n \bm b_2$,  $\sigma=\uparrow\downarrow $, and   $\braket{\bm r}{\bm k,\!\bm Q} = {e^{i(\bm k+\bm Q)\cdot\bm r}}/{\sqrt{A}}$ ( $A$ is the area of the 2D system). In this basis, the noninteracting Hamiltonian can be expressed as a $\bm k$-dependent matrix $h(\bm k)$ whose elements are
\begin{equation}
    h_{\bm Q\sigma,\bm Q'\sigma'}(\bm k)=\mel{\bm k,\bm Q,\sigma}{h(\bm r)}{\bm k,\bm Q',\sigma'}.
\end{equation}
The band dispersion $\epsilon_{n\bm k}$ and the Bloch functions $\bphi_{n\bm k} = [\phi_{n\bk\uparrow},\phi_{n\bk\downarrow}]^T$ are then obtained by diagonalizing $h(\bm k)$. All the Bloch bands are doubly degenerate due to the $\calP\calT$ symmetry of the Hamiltonian, where $\mathcal P=e^{-i\pi\hat L_z}$ is the two-dimensional spatial inversion and $\mathcal T=e^{-i\pi\hat s_y}\mathcal K$ ($\mathcal{K}$ is the complex conjugation) is the time reversal operator for spin-$1/2$ systems.  

We now use (\ref{eq:Sigma_sc}) to calculate $\sigma^{\rm sc}_{\mu\nu}(\omega)$ for the noninteracting Fermi gas. For practical calculations, we need to expand the field operator $\hat\bpsi(\bm r)=[\hat\psi_\uparrow(\bm r),\hat\psi_\downarrow(\bm r)]^T$ in the plane wave basis
\begin{align}
    \hat \bpsi(\bm r) = \frac{1}{\sqrt A}\sum_{\bm k\in {\rm{BZ}},\bm Q} e^{i(\bm k+\bm Q)\cdot \bm r}\hat \bc_{\bm k,\bm Q}
\end{align}
where $\hat \bc_{\bm k,\bm Q}=(\hat c_{\bm k,\bm Q,\uparrow},\hat c_{\bm k,\bm Q,\downarrow})^T$.  Defining $\hat \bpsi_{\bm k}=\big(\hat \bc_{\bm k,\bm Q_1}^T, \hat \bc_{\bm k,\bm Q_2}^T,...\big)^T$, the noninteracting Hamiltonian can be written as 
\begin{align}
\hat h = \sum_{\bm k\in \rm{BZ}} \hat \bpsi_{\bm k}^\dagger h(\bm k) \hat \bpsi_{\bm k},
\end{align}
and the spin density and charge current operators are 
\begin{align}
    \hat \rho^{\rm s}(\bm q)&=\int d\bm r \hat\bpsi^\dagger(\bm r)s_z\hat\bpsi(\bm r)e^{-i\bm q\cdot\bm r} \nn \\
    &=\sum_{\bm k\in\rm{BZ}}\hat \bpsi_{\bm k}^\dagger S_z \hat \bpsi_{\bm k+\bm q},\label{eq:rhosq}\\
    \hat j_{\nu}^{\rm c}(\bm q)&=\frac{1}{2mi}\int d\bm r [\hat\bpsi^\dagger(\bm r)\partial_\nu\hat\bpsi(\bm r)-h.c.]e^{-i\bm q\cdot\bm r} \nn \\
    &=\sum_{\bm k\in\rm{BZ}}\hat\bpsi_{\bm k}^\dagger\partial_{\nu}h(\bm k\!+\!\bm q/2)\hat\bpsi_{\bm k+\bm q},\label{eq:jnuq}
\end{align}
where $S_z=I_{N_{\bm Q}}\otimes s_z$, $N_{\bm Q}$ is the number of $\bm Q$ index of the plane basis and $I_N$ is a $N\times N$ identity matrix. Substituting \eqref{eq:rhosq} and \eqref{eq:jnuq} into Eq.~(\ref{chi_rhoJ1}) of the main text, we find
\begin{align}
    \chi_{\rho^{\rm{s}},j_\nu^{\rm{c}}}(\bm q,\tau) &= -\sum_{\bm k,\bm k'\in\rm{BZ}}\langle\mathscr{T}\hat\bpsi_{\bm k}^\dagger(\tau) S_z \hat\bpsi_{\bm k+\bm q}(\tau) \nn \\
    &\times\hat\bpsi_{\bm k'}^\dagger(0)\partial_{\nu}h(\bm k'\!-\!\bm q/2)\hat\bpsi_{\bm k'-\bm q}(0)\rangle.
\end{align}
Considering the Fourier transform and using the Wick theorem, we have
\begin{align}
   \chi_{\rho^{\rm s},j_\nu^{\rm c}}(\bm q,i\Omega_m) 
    =& \int_0^\beta d\tau \,e^{i\Omega_m\tau} \chi_{\rho^{\rm s},j_\nu^{\rm c}}(\bm q,\tau) \notag \\
    =&  \frac{1}{\beta}\sum_{\bm k\in{\rm{BZ}},n}\tr \left [ S_z G(\bm k+\bm q,i\omega_n+i\Omega_m) \right.\notag \\
    &\times \left.\partial_\nu h(\bm k+\bm q/2)G(\bm k,i\omega_n) \right ], \label{eq:Xw1}
\end{align}
where $\Omega_m=2m\pi/\beta$ and $\omega_n=(2n+1)\pi/\beta$ are the bosonic and fermionic Matsubara frequencies, respectively. Here
\begin{align}
G(\bm k,\tau)&=-\expval*{\mathscr T\hat\bpsi_{\bm k}(\tau)\hat\bpsi_{\bm k}^\dagger(0)} \nn \\
&=\frac{1}{\beta}\sum_n G(\bm k,i\omega_n)e^{-i\omega_n\tau}
\end{align}
is the imaginary time Green's function and its Fourier transform is given by
\begin{align}
    G(\bm k,i\omega_n) & =\int_{0}^{\beta}d\tau\, G(\bm k,\tau)e^{i\omega_n\tau} \nn \\
    & = [i\omega_n-h(\bm k)]^{-1} \nn \\
    &=\sum_n\frac{\ketbra{\phi_{n\bm k}}{\phi_{n\bm k}}}{i\omega_n-\epsilon_{n\bm k}}. \label{eq:Green1}
\end{align}
 Substituting \eqref{eq:Green1} into \eqref{eq:Xw1}, we find
\begin{align}
   & \chi_{\rho^{\rm{s}},j_\nu^{\rm{c}}}(\bm q,i\Omega_m) = \frac{1}{\beta}\sum_{\bm k\in{\rm{BZ}},n,ll'} \nn \\
   &\frac{\bra{\phi_{l\bm k}} S_z\ketbra*{\phi_{l'\bm k+\bm q}}\partial_\nu h(\bm k+\bm q/2)\ket{\phi_{l\bm k}}}{(i\omega_n+i\Omega_m-\epsilon_{l'\bm k+\bm q})(i\omega_n-\epsilon_{l\bm k})}. \label{eq:Xw2}
\end{align}
Summing over the Matsubara frequency $i\omega_n$ and performing the analytic continuation $i\Omega_m \rightarrow \omega+i\eta$, \eqref{eq:Xw2} becomes
\begin{align}
  &  \chi_{\rho^{\rm s},j_\nu^{\rm c}}(\bm q,\omega) = \!\sum_{\bm k\in{\rm{BZ}},ll'}\!\frac{f(\epsilon_{l\bm k})-f(\epsilon_{l'\bm k+\bm q})}{\omega+i\eta+\epsilon_{l\bm k}-\epsilon_{l'\bm k+\bm q}} \nn \\
   & \times \bra{\phi_{l\bm k}}S_z\ketbra*{\phi_{l'\bm k+\bm q}}\partial_\nu h(\bm k+{\bm q}/{2})\ket{\phi_{l,\bm k}}
\end{align}
where $\eta$ is an infinitesimal number but is taken to be a finite spectral broadening parameter later in the plots, $f(\epsilon_{l,\bm k})=\frac{1}{e^{\beta(\epsilon_{l,\bm k}-\mu_F)}+1}$ is the Fermi distribution function and $\mu_F$ is the chemical potential. Finally, the transport tensor \eqref{eq:Sigma_sc} can be calculated numerically by the central difference, i.e.,
\begin{equation}
    \sigma_{\mu\nu}^{\rm{sc}}(\omega)\approx i\frac{\chi_{\rho^{\rm s},j_\nu^{\rm c}}(\bm \delta_\mu,\omega)-\chi_{\rho^{\rm s},j_\nu^{\rm c}}(-\bm\delta_\mu,\omega)}{2|\bm \delta_\mu|},\label{eq:FermiSC}
\end{equation}
where momentum $\bm \delta_\mu$ only contains $\mu$-component and satisfies $|\bm \delta_\mu|/k_L\ll1$. Similarly, we find
\begin{align}
 &   \chi_{j_\nu^{\rm c},\rho^{\rm s}}(\bm q,\omega) =\! \sum_{\bm k\in{\rm{BZ}},ll'}\!\frac {f(\epsilon_{l\bm k})-f(\epsilon_{l'\bm k+\bm q})}{\omega+i\eta+\epsilon_{l\bm k}-\epsilon_{l'\bm k+\bm q}} \nn \\
   & \times \bra{\phi_{l\bm k}}\partial_\nu  h(\bm k+{\bm q}/{2})  \ketbra*{\phi_{l'\bm k+\bm q}}S_z\ket{\phi_{l\bm k}}
\end{align}
from which we can calculate $\sigma_{\nu\mu}^{\rm{cs}}(\omega)$ again by the central difference
\begin{equation}
    \sigma_{\mu\nu}^{\rm{cs}}(\omega)\approx i\frac{\chi_{j_\nu^{\rm c},\rho^{\rm s}}(\bm \delta_\mu,\omega)-\chi_{j_\nu^{\rm c},\rho^{\rm s}}(-\bm\delta_\mu,\omega)}{2|\bm \delta_\mu|}.\label{eq:FermiCS}
\end{equation}

\renewcommand{\theequation}{D\arabic{equation}}
\setcounter{equation}{0}  
\renewcommand{\thefigure}{D\arabic{figure}}
\setcounter{figure}{0}  
\section{Calculation of the charge-spin transport coefficients for the Bose gas }
\label{tcbose}
In this appendix, we provide more details on the ground state symmetry properties of the Bose gas in phase I and II and the calculation of the transport tensors within the Bogoliubov framework. Although some of the information on the ground state symmetry and excitation spectrum can be found in Refs.~\cite{Huang2021, Chen2023, Tang2025}, we sum them up  here for the sake of completeness.
\subsection{Symmetry properties of the meanfield ground states}
\begin{figure*}[tb]
\begin{centering}
\includegraphics[width=15cm]{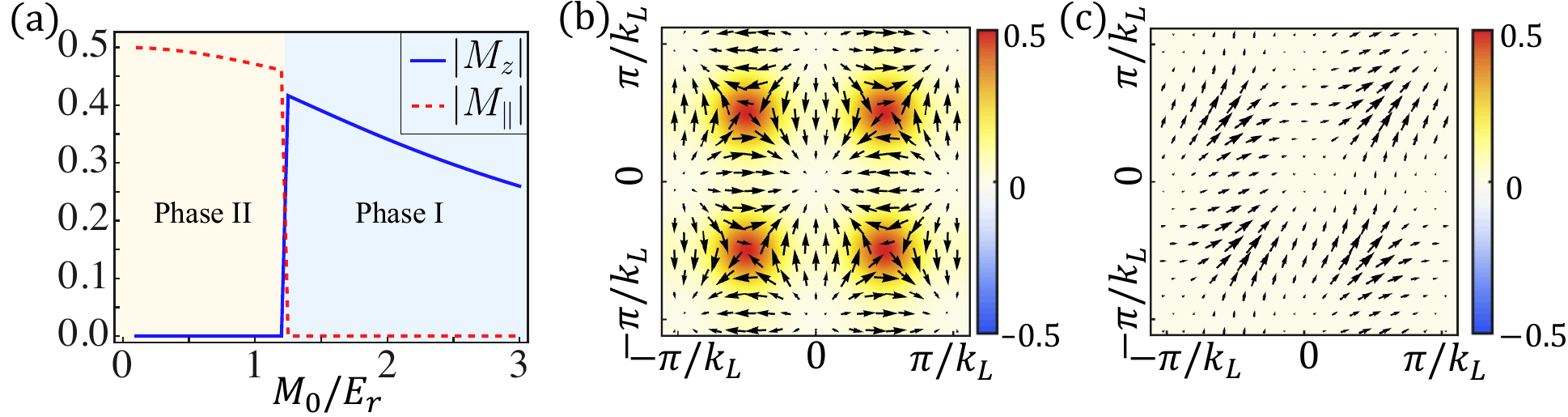}
\par\end{centering}
\caption{(a) As the spin-orbit coupling strength $M_0$ decreases, the Bose gas undergoes a phase transition from the perpendicular magnetizatoin phase (phae I) to the in-plane magnetization phase (phase II). (b) Magnetization density $\bm m(\br)$ of $\bPhi_+$ mode in phase I with $M_0=2.0E_r$. The black arrows denote the $x$- and $y$-components of magnetization density and the background color represents the $z$-component. (c) Magnetization density $\bm m(\br)$ of $\bPhi_1$ mode in phase II with $M_0=1.0E_r$. The rest of the parameters for these three plots are $ V_0=4.0E_r$, $\rho g_{\uparrow\uparrow}=0.25E_r$ and $\rho g_{\uparrow\downarrow}=0.2E_r$.}
\label{fig:phasediagram}
\end{figure*}

At zero temperature, the condensate wave function $\expval*{\hat\bpsi(\bm r)} = [\expval*{\hat\psi_\uparrow(\bm r)} ,\expval*{\hat\psi_\downarrow(\bm r)} ]^T$ can be calculated by minimizing the Gross–Pitaevskii (GP) energy functional $\mathcal{E}(\expval*{\hat\psi_\sigma(\bm r)},\expval*{\hat\psi^\dag_\sigma(\bm r)})$, which can be obtained by simply replacing $\hat \psi_\sigma$ and $\hat \psi^\dag_\sigma$ by $\expval*{\hat\psi_\sigma(\bm r)}$ and $\expval*{\hat\psi^\dag_\sigma(\bm r)}$ respectively in Eq.~(\ref{MBH}). Explicitly, we expand $\expval*{\hat\bpsi(\bm r)}$ in terms of the Bloch functions at the $\Gamma$ point
\begin{equation}
    \expval*{\hat\bpsi(\bm r)}=\bPhi(\bm r)=\sqrt{N}\sum_n c_n\bphi_{n0}(\bm r),
\end{equation}
where $N$ is the atom number of condensate. The coefficients $\{c_n\}$ can be determined by minimizing $\mathcal{E}(\{c_n\})$ and the chemical potential $\mu_B$ is given by $\partial\mathcal{E}/\partial N$. 

Two superfluid phases can be found as the SOC strength $M_0$ is varied, i.e., the perpendicular magnetization phase (phase I) with magnetization along the $z$-direction, and the in-plane magnetization phase (phase II) with magnetization lying in the $xy$-plane \cite{Chen2023}. Here the magnetization $\bM$ and the magnetization density $\bm m(\br)$ are defined as
\begin{align}
\bM = \frac{1}{N}\int d\br\, \bPhi^\dagger(\br) \bs \bPhi(\br) = \int d\br\, \bm m (\br) ,
\end{align}
where $\bs = (s_x,s_y,s_z)$.  The Bose gas in phase I spontaneously selects one of the two degenerate condensate modes $\bPhi_{\pm}$, distinguished by the magnetization $\bm M = |M_z|\big(0,0, \pm 1 \big)$. In phase II, it selects one of the four degenerate modes $\bPhi_l$ distinguished by magnetization $\bm M=|M_\parallel|\big(\cos\frac{(2l-1)\pi}{4},\sin\frac{(2l-1)\pi}{4},0\big)$ ($l=1,...,4.$) where $M_\parallel=\sqrt{M_x^2+M_y^2}$.  We show $|M_z|$ and $M_\parallel$ as a function of $M_0$ in Fig.~\ref{fig:phasediagram}(a). In addition, we also show the magnetization densities of the condensate modes in phase I and II in Fig.~\ref{fig:phasediagram}(b) and (c).  

Now we perform symmetry analysis on the condensate modes in phase I and II and determine what $\mathcal{O}\calT$ symmetries they preserve.  Since the Hamiltonian has $\calP\calT$ symmetry, there are at least two degenerate condensate wave functions $\bPhi'(\bm r)$ and $\bPhi''(\bm r)=\mathcal{PT}\bPhi'(\bm r)$ for the Bose gas in either of the phases. The two states form a basis for a 2D irreducible representation $M^{(\tilde g)}$ of $\tilde D_4$ group, where 
\begin{equation}
\begin{pmatrix}
    M^{(\tilde g)}_{11} & M^{(\tilde g)}_{12} \\
    M^{(\tilde g)}_{21} &M^{(\tilde g)}_{22}
\end{pmatrix} = \begin{pmatrix}
    \la \bPhi'|\tilde g|\bPhi' \ra & \la \bPhi'|\tilde g|\bPhi'' \ra\\
  \la \bPhi''|\tilde g|\bPhi' \ra & \la \bPhi''|\tilde g|\bPhi'' \ra
\end{pmatrix} ,  \forall \tilde g \in \tilde D_4.
\end{equation}
According to Fig.~\ref{fig:phasediagram}(b), the condensate in phase I breaks $C_{2\bm \tau_n}$ symmetries but preserves $C_{4z}^n$ symmetries. So taking $\bPhi' = \bPhi_+$ we can infer that $ C_{4z}^n \bPhi' = e^{i\theta_{z,n}}\bPhi'$ and
\begin{equation}
    C_{2\bm \tau_n} \bPhi'=M_{11}^{(C_{2\bm \tau_n})}\bPhi'+M_{12}^{(C_{2\bm \tau_n})}\bPhi''. \label{eq:snPhi1}
\end{equation}
Noting that in two dimensions $C_{4z}^2 C_{2\bm \tau_n}C_{4z}^{2\dagger} = -C_{2\bm \tau_n}$, we have 
\begin{align}
    M_{11}^{(C_{2\bm \tau_n})}&=\mel{\bPhi'}{C_{2\bm \tau_n}}{\bPhi'} \nn \\ &=\mel{\bPhi'}{C_{4z}^{2\dagger}C_{4z}^2 C_{2\bm \tau_n} C_{4z}^{2\dagger}C_{4z}^2}{\bPhi'} \nn \\ &=-\mel{\bPhi'}{C_{2\bm \tau_n}}{\bPhi'}\nn \\ 
    &=0. \label{eq:Phis2Phi}
\end{align}
Combining \eqref{eq:snPhi1} and \eqref{eq:Phis2Phi}, we then find that $C_{2\bm \tau_n}\bPhi'=e^{i\theta_{\tau_n}}\bPhi''$. On the other hand, $\bPhi'' = \mathcal{PT} \bPhi'$ and thus $C_{2\bm \tau_n}\mathcal{PT}\bPhi'=-e^{-i\theta_{\tau_n}}\bPhi'$ which means that $C_{2\bm \tau_n}\mathcal{PT}$ is a symmetry operation of the state $\bPhi_+$ in phase I. Although there is no restriction for $\boldsymbol\tau_n$ here, there are only two distinctive $\mathcal{O}$ operations that satisfy the conditions in Eq.~(\ref{O}) of the main text, i.e., $ C_{2\bm \tau_2}\mathcal{P}=e^{-i\pi(\hat L_x+\hat s_y)}$ and $ C_{2\bm \tau_4}\mathcal{P}=e^{i\pi(\hat L_y+\hat s_x)}$.

Next, we consider the $\bPhi_1$ mode in phase II (i.e., taking $\bPhi'=\bPhi_1$), which breaks $C_{4z}^n$ symmetries but preserves $ C_{2\bm \tau_1}$ symmetry [see Fig.~\ref{fig:phasediagram}(b)]. Thus we can infer that $  C_{2\bm \tau_1} \bPhi' = e^{i\theta_{\tau_1}}\bPhi'$ and
\begin{equation}
    C_{4z}^2 \bPhi'=M_{11}^{(C_{4z}^2)}\bPhi'+M_{12}^{(C_{4z}^2)}\bPhi''. \label{eq:rnPhi}
\end{equation}
Noting that in two dimensions $ C_{2\bm \tau_1} C_{4z}^2 C_{2\bm \tau_1}^\dagger = -C_{4z}^2$, we have 
\begin{align}
    M_{11}^{(C_{4z}^2)}&=\mel{\Phi'}{C_{4z}^2}{\Phi'} \nn \\
    &=\mel{\Phi'}{ C_{2\bm \tau_1}^\dagger C_{2\bm \tau_1}C_{4z}^2 C_{2\bm \tau_1}^\dagger C_{2\bm \tau_1}}{\Phi'} \nn \\ &=-\mel{\Phi'}{C_{4z}^2}{\Phi'} \nn \\
    & =0. \label{eq:Phir2Phi}
\end{align}
Combining \eqref{eq:rnPhi} and \eqref{eq:Phir2Phi}, we find $C_{4z}^2\bPhi'=e^{i\theta_{z,2}}\bPhi''$. Using $\bPhi'' = \mathcal{PT} \bPhi'$ again, we obtain that $C_{4z}^2\mathcal{PT}\bPhi'=-e^{-i\theta_{z,2}}\bPhi'$ which indicates that $C_{4z}^2\mathcal{PT}$ is a symmetry operation of the state $\bPhi_1$. This is the only $\calO\calT$ symmetry we can identify for $\bPhi_1$ and thus the only $\mathcal{O}$ operator for the condensate in the in-plane magnetized phase is $C_{4z}^2\mathcal{P}=e^{-i\pi \hat s_z}$.

\subsection{Bogoliubov excitations and the charge-spin transport tensors}
\label{sec:Bogoliubov}
Once the condensate wave function $\bPhi$ obatained,  we can substitute $ \hat \psi_\sigma(\bm r)=\Phi_\sigma(\bm r)+\delta\hat\psi_\sigma(\bm r)$
into the Hamiltonian in Eq.~(\ref{MBH}) of the main text to obtain Bogoliubov de Gennes (BdG) Hamiltonian
\begin{equation}
    \hat H_B = \frac 12 \int d\bm r\, \bm {\delta \hat \psi}^\dagger(\bm r) \mathcal{H}(\bm r)\bm{ \delta \hat\psi}(\bm r), \label{eq:Hb1}
\end{equation}
where $\bm{\delta\hat \psi}(\bm r)=[\delta\hat\psi_\uparrow(\bm r),\delta\hat\psi_\downarrow(\bm r),\delta\hat\psi^\dagger_\uparrow(\bm r),\delta\hat\psi^\dagger_\downarrow(\bm r)]^T$ and
\begin{align}
    \mathcal{H}(\bm r)=\mqty(h(\bm r)-\mu_B+\mathcal M(\bm r) & \mathcal N(\bm r)\\ \mathcal N^*(\bm r) & h^*(\bm r)-\mu_B+ \mathcal M^*(\bm r)). \nn
\end{align}
Here 
\begin{align}
	\mathcal{M}(\bm r)=\mqty (2g_{\uparrow\uparrow}|\Phi_{\uparrow}|^{2}+g_{\uparrow\downarrow}|\Phi_{\downarrow}|^{2} & g_{\uparrow\downarrow}\Phi_{\downarrow}^{*}\Phi_{\uparrow} \\
	g_{\uparrow\downarrow}\Phi_{\uparrow}^{*}\Phi_{\downarrow} & 2g_{\downarrow\downarrow}|\Phi_{\downarrow}|^{2}+g_{\uparrow\downarrow}|\Phi_{\uparrow}|^{2} \nn
    )
    \end{align}
   and
    \begin{align}
	\mathcal{N}(\bm r)=\mqty (g_{\uparrow\uparrow}\Phi_{\uparrow}^{2} &g_{\uparrow\downarrow}\Phi_{\downarrow}\Phi_{\uparrow} \\ g_{\uparrow\downarrow}\Phi_{\uparrow}\Phi_{\downarrow} & g_{\downarrow\downarrow}\Phi_{\downarrow}^2). \nn
\end{align}
We solve the BdG Hamiltonian \eqref{eq:Hb1} in plane wave basis. Specifically, we expand operator $\bm{\delta}\hat{\bm \psi}(\bm r)$ as
\begin{equation}
    \bm{\delta\hat \psi(\bm r)}=\frac{1}{\sqrt A}\sum_{\bm k\in{\rm{BZ}},\bm Q} \!(\hat \bc_{\bm k,\bm Q}^T e^{i(\bm k+\bm Q)\cdot\bm r},\hat \bc_{-\bm k,\bm Q}^\dagger e^{i(-\bm k+\bm Q)\cdot\bm r})^T. 
\end{equation}
Defining $\hat \Psi_{\bm k}=\big(\hat\bpsi^T_{\bm k}, \hat\bpsi^\dagger_{-\bm k})^T$ where $\hat \psi_{\bm k}=\big(\hat c_{\bm k,\bm Q_1}^T, \hat c_{\bm k,\bm Q_2}^T,...\big)^T$, the BdG Hamiltonian can be written as 
\begin{align}
\hat H_B = \frac 12 \sum_{\bm k \in\rm{BZ}}\hat \Psi_{\bm k}^\dagger \mathcal H_{\bm k}\hat\Psi_{\bm k},
\end{align}
where
\begin{align}
    \mathcal{H}_{\bm k}=\mqty(h(\bm k)-\mu_B+\mathcal M(\bm k) & \mathcal N(\bm k)\\ \mathcal N^*(-\bm k) & h^*(-\bm k)-\mu_B+ \mathcal M^*(-\bm k)) \nn
\end{align}
with 
\begin{align}
\mathcal{M}_{\bm Q_i,\bm Q_j}(\bm k)&=\frac 1A\int d\bm r\, \mathcal M(\bm r)e^{i(\bm Q_j-\bm Q_i)\cdot \bm r}; \nn \\
\mathcal{N}_{\bm Q_i,\bm Q_j}(\bm k)&=\frac 1A\int d\bm r\, \mathcal N(\bm r)e^{-i(\bm Q_i+\bm Q_j)\cdot \bm r}.
\end{align}
Diagonalizing $\mathcal H_{\bm k}$ by an appropriate paraunitary matrix, we obtain the elementary excitations of the system. 

To calculate the transport tensor \eqref{eq:Sigma_sc},  we turn to the spin-current correlation function in the Bogoliubov framework. The spin density operator and charge current density operator in plane-wave basis and under the Bogoliubov approximation are
\begin{align}
    \hat \rho^{\rm s}(\bm q) &=\sum_{\bm k\in\rm{BZ}}\hat \bpsi_{\bm k}^\dagger S_z \hat \bpsi_{\bm k+\bm q}\nn \\
    &= \frac{1}{2}\sum_{\bm k\in\rm{BZ}}\hat \Psi_{\bm k}^\dagger \mathcal S_z\hat \Psi_{\bm k+\bm q} \nn \\
    &\approx  \Psi_0^\dagger \mathcal S_z\hat \Psi_{\bm q}\label{eq:szq}
    \end{align}
    and
    \begin{align}
    \hat j_{\nu}^{\rm c}(\bm q)&=\sum_{\bm k\in\rm{BZ}}\hat\bpsi_{\bm k}^\dagger\partial_{\nu}h(\bm k\!+\!\frac{\bm q}{2})\hat\bpsi_{\bm k+\bm q} \nn \\&= \frac{1}{2}\sum_{\bm k\in\rm{BZ}}\hat \Psi_{\bm k}^\dagger \mathcal J^{\rm c}_\nu(\bm k\!+\!\frac{\bm q}{2})\hat \Psi_{\bm k+\bm q} \nn \\
    &\approx  \hat\Psi^\dagger_{-\bm q}  \mathcal J^{\rm c}_\nu(-\frac{\bm q}{2})\Psi_0 \label{eq:jnuqb}
\end{align}
respectively, where $\mathcal S_z = I_2\otimes S_z$, $\mathcal J^{\rm c}_\nu(\bm q) = {\rm{diag}}\big(\partial_\nu h(\bm q),\,\partial_\nu h^*(\bm k)|_{\bm k=-\bm q}\big)$ and $\Psi_0 = \big[\langle\hat\bpsi_0^T\rangle, \langle\hat\bpsi^\dagger_0\rangle\big]^T$. Explicitly, we have $\langle\hat\bpsi_0^T\rangle=\big(\langle\hat{\bm c}_{0,\bm Q_1}^T\rangle,\langle\hat{\bm c}_{0,\bm Q_2}^T\rangle,...\big)$ with
\begin{equation}
    \quad\langle\hat{\bm c}_{0,\bm Q_i}\rangle=\frac{1}{\sqrt{A}}\int d\bm r\, \bPhi(\bm r)e^{-i\bm Q_i\cdot \bm r}.
\end{equation}
Using  Eqs.~\eqref{eq:szq} and \eqref{eq:jnuqb}, the temperature spin-current correlation function can be written as
\begin{align}
    \chi_{\rho^{\rm s},j_\nu^{\rm c}}(\bm q,i\Omega_m) = \Psi_0^\dagger \mathcal S_z \mathcal G(\bm q,i\Omega_m) \mathcal J^{\text c}_\nu(\bm q/2)\Psi_0,
\end{align}
where $\mathcal G(\bm q,i\Omega_m)$ is the retarded Green's function of the BdG Hamiltonian
\begin{align}
    \mathcal G(\bm q, i\Omega_m) &= -\int_0^\beta d\tau \la\mathscr T \Psi_{\bm q}(\tau) \Psi_{\bm q}^\dagger(0)]\ra  e^{i\Omega_m \tau} \nn \\
    &= [i\Omega_m \tau_z -\mathcal H_{\bm q}]^{-1}.
\end{align}
Here  $\tau_z=\sigma_z\otimes I_{2N_{\bm Q}}$. Finally, the transport tensor is given by
\begin{align}
    \sigma_{\mu\nu}^{\rm{sc}}(\omega) = & i\partial_{q_\mu} \chi_{\rho^{\rm s},j_\nu^{\rm c}}(\bm q,i\Omega_m\rightarrow\omega+ i\eta)\big|_{\bm q=0} \nn \\
    = & i\Psi_0^\dagger \mathcal S_z \big[\mathcal G(0,\omega)\tau_z \mathcal J^{\rm c}_\mu(0)\mathcal G(0,\omega) \mathcal J^{\rm c}_\nu(0) \nn \\
    &+  \mathcal G(0,\omega) \partial_\mu \mathcal J^{\rm c}_\nu(0)/2\big]\Psi_0
    \label{sigmascB}
\end{align}
where $\partial_{\bm q}\mathcal G = -\mathcal G(\partial_{\bm q}\mathcal G^{-1})\mathcal G$ and $\partial_{\bm q}\mathcal G^{-1}(\bm q) =-\partial_{\bm q}\mathcal H_{\bm q} =-\tau_z \mathcal J^{\text c}_\mu(\bm q)$ have been used. Similarly, we find that the reciprocal tensor $\sigma_{\nu\mu}^{\text{cs}}(\omega)$ is given by
\begin{align}
    \sigma_{\nu\mu}^{\rm{cs}}(\omega) = & i\partial_{q_\mu} \chi_{j_\nu^{\rm c},\rho^{\rm s}}(\bm q,i\Omega_m\rightarrow\omega+ i\eta)\big|_{\bm q=0} \nn \\
    = & i\Psi_0^\dagger \big[\partial_\mu \mathcal J^{\rm c}_\nu(0)\mathcal G(0,\omega)/2  \nn \\
    & +\mathcal J^c_\nu(0)\mathcal G(0,\omega)\tau_z \mathcal J^{\rm c}_\mu(0)\mathcal G(0,\omega) \big ]\mathcal{S}_z\Psi_0.
    \label{sigmacsB}
\end{align}

\bibliography{Onsager_REFS}
\end{document}